\documentclass[aps,pra,10pt,superscriptaddress,eqsecnum,nofootinbib,showpacs,preprintnumbers]{revtex4-2}
\usepackage{setspace} 

\usepackage{amsmath,mathtools}
\usepackage{amssymb}
\usepackage{amsfonts,amsthm,bm}
\usepackage{color,xcolor}
\usepackage[greek,english]{babel}
\usepackage{slashed}
\usepackage{titlesec}
\usepackage{comment}
\usepackage{amsmath,mathtools}
\usepackage{amssymb}
\usepackage{amsfonts,amsthm,bm}
\usepackage{color,xcolor}
\usepackage[greek,english]{babel}
\usepackage{slashed}
\usepackage{titlesec}
\usepackage{comment}
\usepackage{appendix}
\usepackage{graphicx,epsfig}
\graphicspath{{figures/}{./}}
\usepackage{subfigure}
\usepackage{tikz}

\newcommand{\abs}[1]{\left| #1 \right|}

\newcommand{\be}{\begin{eqnarray}}
	\newcommand{\ee}{\end{eqnarray}}
\newcommand{\bea}{\begin{eqnarray}}
	
	\newcommand{\eea}{\end{eqnarray}}

\newcommand{\beq}{\begin{equation}}
	\newcommand{\eeq}{\end{equation}}
\newcommand{\bseq}{\begin{subequations}}
	\newcommand{\eseq}{\end{subequations}}
\let\oldsqrt\sqrt
\def\sqrt{\mathpalette\DHLhksqrt}
\def\DHLhksqrt#1#2{%
\setbox0=\hbox{$#1\oldsqrt{#2\,}$}\dimen0=\ht0
\advance\dimen0-0.2\ht0
\setbox2=\hbox{\vrule height\ht0 depth -\dimen0}%
{\box0\lower0.4pt\box2}}
\usepackage[colorlinks=true,linkcolor=blue,citecolor=red,urlcolor=magenta]{hyperref}

\begin{document}
\preprint{\leftline{KCL-PH-TH/2025-23}}
 
	\title{Dynamical-System analysis of single-axion monodromy inflation with periodically-modulated potentials}
	
	\author{Panagiotis  Dorlis }
	\email{psdorlis0@gmail.com} 
	\affiliation{Physics Division, School of Applied Mathematical and Physical Sciences,
		National Technical University of Athens,  Zografou Campus,
		Athens 15780, Greece.}
	\author{Nick E. Mavromatos}
	\affiliation{Physics Division, School of Applied Mathematical and Physical Sciences,
		National Technical University of Athens,  Zografou Campus,
		Athens 15780, Greece.}
	\affiliation{Theoretical Particle Physics and Cosmology Group, Physics Department, King's College London, Strand, London WC2R 2LS, UK.}

	\author{Sotirios-Neilos Vlachos }
	\email{sovlacho@gmail.com}
	\affiliation{Physics Division, School of Applied Mathematical and Physical Sciences,
		National Technical University of Athens,  Zografou Campus,
		Athens 15780, Greece.}
	
	\author{Makarios Vyros }
	\email{makariosvyr@gmail.com}
	\affiliation{Physics Division, School of Applied Mathematical and Physical Sciences,
		National Technical University of Athens,  Zografou Campus,
		Athens 15780, Greece.}

	\vspace{17.5cm}
	\begin{abstract}
In this work, we study field theoretic systems of a single axion-like field with linear potentials modulated by cosine terms, allegedly induced by non-perturbative instanton configurations. These systems are considered 
in expanding-Universe spacetime backgrounds (of Friedman-Lema$\hat{\rm i}$tre-Robertson-Walker type). Using a dynamical-system approach, we classify the various de-Sitter like (inflationary) vacua from the point of view of their stability, which depend on the values of the model parameters. In this respect, bifurcation points are found to be present for the various models under consideration. Part of the parameter space of the systems under consideration includes the running-vacuum (approximately) linear-axion monodromy potentials, considered in previous works by some of the authors, where inflation is induced by primordial gravitational-wave condensates. A particularly interesting case, corresponding to another part of the parameter space of the models, includes a series of stable de-Sitter vacua, which physically may correspond to a series of successive tunnelings of the system, via say non-perturbative effects, with a decreasing effective cosmological constant. Under certain values of the parameters, these successive tunnelings can reach a Minkowski spacetime, with zero value of the minimum of the axion potential. The situation is not dissimilar to the one of discrete inflation that arguably characterizes some minimal non-critical-string (Liouville) models of cosmology. Finally, for comparison, we also include in this article a dynamical-system study of  standard axion-monodromy-modulated potentials characterizing some string/brane-compactification models of inflation.
   
	\end{abstract}
	\vspace{3.5cm}

	\maketitle
	
	\flushbottom
	
	\tableofcontents

\section{Introduction}

The Dynamical-Systems approach to Cosmology~\cite{Bahamonde:2017ize} proves to be a powerful tool to study early Universe cosmology, and in particular inflation~\cite{inflationaris} and its attractor behavior~\cite{Hao:2003aa,Faraoni:2012bf,Bouhmadi-Lopez:2016dzw,Bahamonde:2017ize,Hossain:2023lxs}. In a recent paper~\cite{Dorlis:2024yqw}, this method has been applied to a 
model of inflation in the context of a string-inspired~\cite{bms,bms2,ms1,ms2,Mavromatos:2022xdo} Chern-Simons (CS) anomalous gravity~\cite{Jackiw:2003pm,Alexander:2009tp}. In this scenario, the inflationary epoch is driven by the formation of a primordial-gravitational-wave (GW) induced condensate of the gravitational CS anomaly terms, which couple to a massless axion field $b(x)$, the so-called string-model independent axion in the standard string terminology~\cite{svrcek} (in string theory there also other axions $a^I(x), I=1,2, \dots N_a$ coming from string compactification, hence dependent on the details of the underlying microscopic string model). 

The condensate leads to a running-vacuum-model (RVM) type cosmology~\cite{SolaPeracaula:2022hpd,Moreno-Pulido:2023ryo}. The inflation in this stringy RVM (StRVM) cosmology is essentially induced by the linear-axion potential due to the formation of the CS condensate:\footnote{The StRVM cosmology can also provide a natural explanation on the alleviation of the Hubble and growth-of-structure tensions in the current era~\cite{Gomez-Valent:2023hov}, in agreement with the spirit of the generic RVM cosmologies~\cite{SolaPeracaula:2023swx}.}
\begin{align}\label{Vlin}
V(b) = \Lambda_0^3 \, b(x)\,,    
\end{align}
where $b(x)$, of mass dimension +1, is a canonically normalized axion field ({\it i.e.} has a standard kinetic term $\frac{1}{2}\sqrt  {-g}\, \partial_\mu b \, \partial^\mu b$ in the (3+1)-dimensional effective Lagrangian density after compactification, on a four-dimensional spacetime background with metric $g_{\mu\nu}$, $\mu, \nu = 0, \dots 3$). The quantity $\Lambda_0^3$, where $\Lambda_0$ is a mass scale, is the CS condensate~\cite{bms,Dorlis:2024yqw}, which during inflation (suffix $``I''$) assumes the approximately constant form:
\begin{align}\label{l0}
\Lambda_0^3 = A\, {\rm Re} \left \langle R_{CS} \right \rangle 
\equiv  A\, {\rm Re}\,\left\langle \frac{1}{2} R^\mu_{\,\,\nu\rho\sigma}\, {\widetilde R}^{\nu \,\,\,\rho\sigma}_{\,\,\mu} \right \rangle 
= - \frac{A^2}{\pi^2}\, \frac{{\dot b}_I}{M_{\rm Pl}}\, \Big(\frac{H_I}{M_{\rm Pl}}\Big)^3 \, \mu^4\,,     
\end{align}
where $\widetilde{(\dots)}$ denotes the dual of the Riemann tensor in (3+1)-dimensional spacetime, $H_I$ is the Hubble parameter during inflation, for which phenomenologically we have~\cite{Planck}\,:
\begin{align}\label{MIMPl}
H_I \lesssim 2.5 \times 10^{-5}\, M_{\rm Pl}\,, 
\end{align}
where $\langle \dots \rangle$ denotes a condensate of GW modes, whose momenta are cut off in the ultraviolet (UV) at an energy scale $\mu$, and $A$ is the coefficient of the CS gravity, which in string theory is given by~\cite{kaloper}:
\begin{align}\label{Adefstr}
A = \sqrt{\frac{2}{3}}\, \frac{1}{48}\, \mathfrak{a}^\prime \, M_{\rm Pl}\,,
\end{align}
which has mass dimension $-1$. Above, $\mathfrak{a}^\prime = M_s^{-2}$ is the Regge slope, with $M_s$ the string mass scale (which is, in general, different from the reduced Planck mass $M_{\rm Pl}$). In the context of string-inspired effective low-energy theories, such as the StRVM~\cite{bms,ms1,Dorlis:2024yqw}, one has: 
\begin{align}\label{msmp}
\mu=M_s\,. 
\end{align}
It should be remarked at this stage that, to ensure a sub-Planckian string scale, $M_s \le M_{\rm Pl}$, thus satisfying the transplanckian-censorship conjecture~\cite{trans}, we need to consider a macroscopic number $\mathcal N_I$ of sources of GW, which enhances the condensate in a way proportional to this number~\cite{Mavromatos:2022xdo,Dorlis:2024yqw}. 
From a physical point of view, the presence of a large number $\mathcal N_I$ compared, say, to the corresponding number of GW sources during the stiff era preceding the inflationary epoch in the StRVM~\cite{bms,ms1,ms2}, helps ensuring~\cite{Dorlis:2024yqw} a constant CS condensate in the transition from the stiff to the inflationary era. 

The alert reader should notice that 
it is the real part (denoted in \eqref{l0} by ``Re'') of the CS condensate that determines the linear axion potential \eqref{Vlin}, which we use in the dynamical system analysis~\cite{Dorlis:2024yqw}. Indeed, the existence of imaginary parts in the condensate, which arise in the computation using weak graviton approximation, indicate metastability of the inflationary era, and they help in determining the duration of inflation. The dynamical system analysis, using appropriate boundary conditions, turns out to be remarkably consistent with the condensate analysis~\cite{Dorlis:2024uei} insofar as the duration of inflation is concerned between these two approaches. By requiring consistency with the cosmological data~\cite{Planck}, pointing towards lifetimes of ${\mathcal O}(50-60)$ e-foldings, one obtains the condition~\cite{Dorlis:2024uei}
\begin{align}\label{msmpo}
\frac{M_s}{M_{\rm Pl}} = \mathcal O(10^{-1}) \,.
\end{align}
Metastability is also crucial for the embedding of the StRVM into a consistent microscopic string theory, thus avoiding the swampland~\cite{swamp1,swamp2,swamp3,swamp4}.  

The linear potentials \eqref{Vlin}, \eqref{l0}, are valid for the entire range of values of the axion field $b(x) \in \mathbb R$, and not only for large values (compared to a fundamental mass scale), as is the case of the so-called axion-monodromy potentials that characterized cosmological models from string/brane theory~\cite{silver}, after compactification to (3+1) spacetime dimensions. For these latter systems, on assuming a single dominant compactification-axion species for concreteness and brevity, the potential of the (dimensionless) axion ($a$) field has the form: 
\begin{align}\label{Vmonodr}
V(a)=\frac{\epsilon}{g_s(2 \pi)^5 \mathfrak{a}^{\prime 2}} \sqrt{\ell^4+a(x)^2}\,, 
\end{align}
where $\epsilon$ is a constant related to the appropriate brane compactification in the model, $\ell > 1$ is  a dimensionless positive number 
that is associated with the size of the compact dimensions ({\it e.g.} a five brane wrapped around a two cycle $\Sigma^{(2)}$~\cite{silver}), taken for simplicity here to have a typical radius of order $\ell \sqrt{\mathfrak{a}^\prime}$, and $g_s$ is the string coupling. This potential leads to a linear axion potential for large values of the (dimensionless) axion field $|a(x)| \gg \ell^2$, 
\begin{align}\label{Vlincomp}
V(\phi_a) \simeq \Lambda_a^3 \, |\phi_a(x)|\,, \qquad \Lambda_a^3 \equiv \frac{\epsilon}{g_s\, (2\pi)^5} \, \frac{M_s^4}{f_a} \equiv \frac{\Lambda_2^4}{f_a}\,, 
\end{align}
where $f_a$ is the compactification-axion-$a$ coupling, of mass-dimension $+1$, corresponding to the canonically-normalized axion field $\phi_a$, of mass dimension $+1$, obtained from the dimensionless field $a$ by the redefinition $f_a \, a = \phi_a$. On comparing \eqref{Vlin} with \eqref{Vlincomp}, we observe that, in the former case the linear axion behavior is exact, due to the exact one-loop nature of the CS anomaly term, 
while in the latter case, the linear behavior of the axion potential is approximate, valid only for the regime of relatively large fields, compared to an appropriate fundamental mass scale. Moreover, in this case the potential is symmetric under the field reflection $a(x) \to - a(x)$, in contrast to the condensate-inflation case \eqref{Vlin}, in which the axion field has different behavior on negative values. 

In \cite{Dorlis:2024uei}, we consider the effects of gauge-field instantons in the target-space of the microscopic string model underlying the StRVM effective low-energy cosmology of \cite{bms,ms1}. These provide periodic modulation of the linear-axion potential \eqref{Vlin}:
\begin{align}\label{permod}
 V_{\rm inst} =\Lambda_0^3 \, b(x) + \Lambda_1^4 \, \cos \Big(\frac{b(x)}{f_b}\Big)\,,   
\end{align}
where $\Lambda_1 = \xi M_s^4 \exp \Big(-S_{\rm inst}\Big)$ is the scale of the instantons, $\xi$ is some phenomenological number, and $S_{\rm inst} \gg 1$ is the large Euclidean instanton action in (3+1)-dimensional spacetime after string compactification in the StRVM. The parameter $f_b > 0$, with mass dimension $+1$ is the coupling of the string-model-independent axion field $b(x)$, which in the StRVM assumes the value: 
\begin{align}\label{fb}
    f_b = \frac{1}{16\, \pi^2\, A} = \sqrt{\frac{3}{2}} \, \frac{3}{\pi^2} \, \frac{M_s^2}{M_{\rm Pl}} \simeq 0.37\,\frac{M_s^2}{M_{\rm Pl}}\,, 
\end{align}
given that the axion coupling $f_b$ is defined as the inverse of the coefficient in 
the (3+1)-dimensional action term of the gauge anomaly $\frac{1}{f_b}\,  \frac{1}{16\pi^2} \, \int d^4 x\, {\rm Tr}\, \mathbf F_{\mu\nu} \, \widetilde{{\mathbf F}}^{\mu\nu}$, where the quantity under integration is associated with the Pontryagin index~\cite{Eguchi:1980jx}, and equals $16\pi^2 \, n$, where $n \in \mathbb Z$ is an integer. This explains the periodic modulation \eqref{permod} of the axion potential due to gauge instanton configurations. 

In \cite{Dorlis:2024uei}, the effect of instantons with actions of $\mathcal O(5)$, and $\xi = \mathcal O(1)$ in the expression for the scale $\Lambda_1$, which were compatible with strongly coupled Yang-Mills couplings $g_{\rm YM}$, corresponding to fine structure constant $\frac{g_{\rm YM}^2}{4\pi} = \mathcal O(1)$, has been shown to amount to a permil shift of the values of the corresponding slow-roll inflationary parameters~\cite{inflationaris} of the axion potential \eqref{permod}. This results in a  shift of the value of the inflationary spectral index $n_s$, so that it is now closer to the central value inferred by the cosmological measurements~\cite{Planck}. This case, corresponds to the following scale hierarchy
\begin{align}\label{scalehierarchy}
    \Lambda_0 \gg \Lambda_1 \,.
\end{align}
In such a limiting case, the classification of the inflationary points by means of the dynamical system analysis can be provided exclusively by the linear term in the potential \eqref{permod}, for all practical purposes, in the way studied in \cite{Dorlis:2024uei,Dorlis:2024yqw}. 

Periodic modulations due to world-sheet instantons can also affect the compactification-axion monodromy potential $\phi_a(x)$ of ref.~\cite{silver}, mentioned above, {\it cf.} \eqref{Vmonodr}, which assumes the form:
\begin{align}\label{Vmonodr2}
  V(\phi_a)^{\rm ws~inst} = \frac{\epsilon\, \ell^2}{g_s\, (2\pi)^5} \,  M_s^4 \sqrt{1 + \Big(\frac{\phi_a(x)}{\ell^2\, f_a}\Big)^2} + \Lambda_{\rm ws}^4 \, \cos \Big(\frac{\phi_a(x)}{2 \pi f_a}\Big)\,. 
\end{align}
In \eqref{Vmonodr2}, the scale $\Lambda_{\rm ws}$ pertains to the world-sheet instantons of the underlying brane/string model~\cite{silver}. In the large-field limit, $\phi_a \gg \ell^2 \, f_a$, one obtains from \eqref{Vmonodr2} 
a periodically modulated potential \eqref{Vlincomp}, that is, 
\begin{align}
\lim_{\phi_a \gg \ell^2\, f_a} V(\phi_a) = \Lambda_a^3\,|\phi_a(x) | + \Lambda_{\rm ws}^4 \, \cos \Big(\frac{\phi_a(x)}{2 \pi f_a}\Big)\,.
\end{align}

In this work we would like to go beyond the cases examined above, and consider periodically modulated potentials of the form \eqref{permod} or \eqref{Vmonodr2}, but keeping the hierarchy of scales $\Big(\Lambda_0, \, \, \Lambda_1 \Big)$ 
and $\Big( \sqrt{\ell}\,\Lambda_2, \, \, \Lambda_{\rm ws} \Big)$, respectively for the two cases, arbitrary. Such a general case could characterize generic CS gravity models~\cite{Jackiw:2003pm,Alexander:2009tp}, where the coupling $A$ is no longer given by \eqref{Adefstr}, but considered as a phenomenological parameter. Similarly, \eqref{msmp} is no longer valid in general, since the underlying theory might not be strings, in which case a natural value of the graviton UV cutoff $\mu$ should be $M_{\rm Pl}$. With generic values of the CS parameter $A$, the hierarchy between the scales $\Lambda_0$ and $\Lambda_1$
in \eqref{permod} is kept free, and one should examine cases in which the magnitude of the scale of the periodic modulation terms is larger, of similar order, or smaller than that of the linear terms. Similar general considerations can also be adopted for the string/brane compactification-axion case \eqref{Vmonodr2}, studied in \cite{silver}. 

Our aim is to use dynamical systems analysis for such generalized axion cosmologies, and study the emergence and (meta)stability of inflationary de-Sitter like points, following our previous study in \cite{Dorlis:2024yqw}. 
We commence our analysis with the study of potentials of the form \eqref{permod} for the string independent axion $b(x)$. 
Then, we perform the analysis for the string/membrane-inspired model for the compactification axion $a(x)$, with potential \eqref{Vmonodr2}. We shall seek fixed points in the dynamical-system phase space and examine their stability, generalizing the analysis in \cite{Dorlis:2024yqw}. We are particularly interested in points corresponding to metastable de Sitter vacua, which are in principle embeddable in microscopic UV complete quantum gravity models, and are associated with phenomenologically realistic inflation life times of ${\mathcal O}(50-60)$ e-foldings~\cite{Planck,inflationaris},
but we shall also demonstrate the presence of  saddle points, as well as points  corresponding to eternal de-Sitter spacetimes (the latter are excluded as consistent quantum-gravity theories according to the swampland criteria~\cite{swamp1,swamp2,swamp3,swamp4}).
The presence of periodic modulations in the linear (or square-root type, in the string/brane compactification case) potentials, leads, under some conditions, to an interesting sequence of fixed points, such that the corresponding cosmology passes through a series of tunnelings before settling down to a stable vacuum.
We therefore hope that our classification and conclusions in the current article will prove to be of complementary use in studies of inflationary models, even beyond the axionic ones. 

The structure of the article is the following:
in the next section, \ref{sec:dynsys}, we discuss the general formalism to be used in the dynamical system analysis of an axion cosmology with a potential $V(b)$ given by \eqref{permod}. We parametrize the relative scale hierarchy between $\Lambda_0$ and $\Lambda_1$ by using a suitable set of parameters
$\gamma$ and $\delta$ (see \eqref{gd} below) and $\widetilde{\lambda} $ (see \eqref{lambda0}). In section \ref{sec:3}, we study the case $\widetilde{\lambda} \geq 0$ and $\gamma < 0$ and its 
stability. In section \ref{sec:5}, we study the classification of the models corresponding to different values of the ratio $\gamma/\delta$. Our aim is to determine phenomenologically interesting models of metastable inflation, with duration $50-60$ e-foldings, or points in the respective phase space corresponding to eternal de-Sitter spacetimes. In section \ref{sec:five}, we study the case $\widetilde{\lambda} < 0$ and $\gamma < 0$ and its stability. In section \ref{sec:6}, we repeat the above analysis for some cases in the string/brane-inspired model of \cite{silver}, described by a 
compactification-axion potential of the form \eqref{Vmonodr2}, where now the periodic modulations come from world-sheet instantons of the underlying string/brane microscopic theory. Finally, section \ref{sec:7} contains our conclusions. Some technical aspects of our work, concerning the pure-cosine potential case, as well as some mathematical aspects of stability criteria, including the use of center manifold  
theory~\cite{Bahamonde:2017ize,boehmer2010,Boehmer:2011tp} to study the stability of two special cases (the case with $\abs{\gamma/\delta} =1$, characterized by a sequence of saddle fixed-points, and the infinite-compactification radius (formal) case of the string/brane-inspired model~\cite{silver}), are given in Appendices \ref{app2}, \ref{appB} and \ref{app3}, respectively.

\section{Dynamical Systems Analysis of (Pseudo)Scalar Field Cosmology: General Formalism}\label{sec:dynsys}

In this section, we study the dynamics of a (pseudo)scalar field cosmology by considering a minimally coupled (pseudo)scalar field in a Friedmann-Lema$\hat{\rm i}$tre-Robertson-Walker (FLRW) spacetime geometry. Our basic goal here is to translate the Friedmann and Klein-Gordon equations, that govern the cosmological evolution, into the language of dynamical systems~\cite{B_hmer_2016}. Following the analysis presented in \cite{Dorlis:2024yqw}, we consider the general action for a minimally coupled interacting (pseudo)scalar field $b(x)$:
\begin{align}
	S=\int d^4x \,\sqrt{-g}\,  \left[\frac{R}{2\kappa^2}-\frac{1}{2}(\partial_\mu b)(\partial^\mu b) -V(b) \right]\,,
\label{eq:Action_dynamical}
\end{align}
where $V(b)$ denotes a self-interaction potential for the $b(x)$ field. As usual, the gravitational field equations are given by:
\begin{align}
	G_{\mu\nu}=\kappa^2 T^{b}_{\mu\nu}\,,
\label{eom1}
\end{align}
where $G_{\mu\nu} = R_{\mu\nu} - \frac{1}{2} g_{\mu\nu} R$ is the Einstein tensor, and 
\begin{align}\label{stress}
    T^b_{\mu\nu}=\partial_\mu b\partial_\nu b-\frac{1}{2}g_{\mu\nu}(\partial b)^2 - g_{\mu\nu}V(b)
\end{align}
is the stress energy tensor of the (pseudo)scalar field $b$. After varying \eqref{eq:Action_dynamical} with respect to the field $b(x)$, we obtain the Klein-Gordon equation in curved spacetime:
\begin{align}
    \Box b -V_{, b} = 0\,,
\end{align}
where commas denote functional differentiation with respect to the field $b(x)$. We consider a spatially flat FLRW spacetime line element, 
\begin{align}
    ds^2=-dt^2+\alpha^2(t)\delta_{ij}dx^idx^j\,,
\end{align}
with $\alpha (t)$ denoting the scale factor (we take $\alpha_0=1$ for today's value, as usual). With this choice of spacetime, the gravitational equations of motion \eqref{eom1} reduce to the Friedmann equations, that is\,:
\begin{align}
    3H^2 = & \,\kappa^2  \,\left(\frac{\dot{b}^2}{2}+V(b)\right)\,,
    \label{friedmann1}\\ 
    2\dot{H} + 3H^2 = & -\kappa^2\left(\frac{\dot{b}^2}{2}-V(b)\right)  \Rightarrow \dot{H} \,\stackrel{\eqref{friedmann1}}{=}\, -\frac{1}{2}\,\kappa^2 \,\dot{b}^{2}\,,
    \label{friedmann2}
\end{align}
and the Klein-Gordon equation for the scalar field assumes the form:
\begin{align}
    \ddot{b}+3H\dot{b}+V_{,b}=0\,.
    \label{KG_frw}
\end{align}
In our cosmological setting, the energy density and the pressure for the scalar field fluid are given by:
\begin{align}
    \rho_b = \frac{\dot{b}^2}{2}+V(b)\,, \qquad p_b = \frac{\dot{b}^2}{2}-V(b)\,,
\end{align}
where the equation of state (EoS), $p_b=w_b \,\rho_b$, is dynamical for the (pseudo)scalar field and reads:
\begin{align}
w_b=\frac{\frac{\dot{b}^2}{2}-V(b)}{\frac{\dot{b}^2}{2}+V(b)}\,,
    \label{eq_of_State}
\end{align}
which takes on values in the range from $-1$ (dominance of potential energy) to $+1$ (dominance of kinetic energy). The former scenario is physically appealing, as the (pseudo)scalar field can give rise to inflation. We aim to study the cosmological evolution as a dynamical system with the potential for $b(x)$ given by \cite{Dorlis:2024uei}:
\begin{align}\label{pot}
V(b) = \Lambda_{0}^3 \,b  + \Lambda_1^4 \, {\rm cos}\left(\frac{b}{f_b}\right)\,,
\end{align}
with $\Lambda_0,\Lambda_1$ corresponding to two different energy scales introduced to this model. Such potentials have been motivated in the introduction of the current paper.

Before proceeding with the solutions of the dynamical systems that correspond to the equations~(\ref{friedmann1})--(\ref{KG_frw}), it is important to make come comments regarding the (approximate) mapping of the gravitational-anomaly-condensate-induced inflationary scenario of \cite{Dorlis:2024yqw,Dorlis:2024uei}, within the framework of CS gravity~\cite{Jackiw:2003pm}, 
onto an axion-like system with potentials of the form \eqref{pot}. An important feature of such a CS-gravity model is a linear coupling of the axion to the gravitational CS term, which, upon condensation of the anomaly terms, induced by primordial GWs, leads to an axion system with an (approximately) linear potential, plus periodic modulations, of the form \eqref{pot}. 

However, contrary to the simple system \eqref{eq:Action_dynamical}, discussed above, a non-trivial gravitational anomaly, as is the case of 
cosmologies with primordial GWs, leads~\cite{Jackiw:2003pm} to the presence of non-trivial  components of the Cotton tensor, stemming from the gravitational variation of the CS anomaly terms, on the left-hand-sides of Eqs.~\eqref{eom1}. This drastically changes the picture, as far as  the existence of inflationary solutions with $H = H_I \simeq \rm constant \neq 0$, $\dot{b} \simeq \rm constant\neq 0$, is concerned. Indeed, the condensation of the CS anomaly implies the consistency of such solutions, in contrast to the simplified Friedmann equations \eqref{friedmann1}, \eqref{friedmann2}, which necessarily imply $\dot b=0$ during the inflationary era, for which $H = \rm constant$. Nonetheless, as demonstrated in  \cite{Dorlis:2024yqw}, there is a remarkable consistency between the features of CS-condensate-induced inflation and the dynamical-system analysis, based on potentials of the form \eqref{pot}. Indeed, the analysis of \cite{Dorlis:2024yqw}, has demonstrated that, during the phenomenologically relevant~\cite{Planck} inflationary era ({\it cf.} \eqref{MIMPl}), 
determined by an appropriate choice of the initial conditions of the corresponding dynamical system, one has 
$\dot b \sim 10^{-1} H_I \, M_{\rm Pl}$, which, on account of \eqref{friedmann2} and \eqref{MIMPl}, would give, 
\begin{align}\label{Hdot}
\abs{\dot H_I} \sim 5 \times 10^{-3} \Big(\frac{H_I}{M_{\rm Pl}}\Big) \, H_I M_{\rm Pl} \lesssim 1.25 \times 10^{-7} H_I M_{\rm Pl} \ll H_I M_{\rm Pl}\,, 
\end{align}
thus implying that, to an excellent approximation, the slow-roll conditions for inflation are satisfied, 
being correctly captured by the dynamical system of \cite{Dorlis:2024yqw}. Moreover, the latter reproduces all the essential features of the condensate-induced inflation of \cite{bms,ms1,ms2}. This motivates the approximate use of  dynamical systems even for the modulated anomalous-CS-condensate gravity cases, based on the potential \eqref{pot}, examined in this work. 

We are now ready to embark on our analysis of the dynamical system corresponding to the equations~(\ref{friedmann1})--(\ref{KG_frw}), with the potential \eqref{pot}. 
To this end, we first note that the derivatives of the potential with respect to the $b(x)$-field are given by:
\begin{align}\label{pot11}
V_{, b} (b) = \Lambda_{0}^3  - \frac{ \Lambda_1^4}{f_b} \, {\rm sin}\left(\frac{b}{f_b}\right), \ \ \ \ \ V_{, b b} (b) =  - \frac{ \Lambda_1^4}{f_b ^2} \, {\rm cos}\left(\frac{b}{f_b}\right) \,.
\end{align} 
We may now introduce the dimensionless \textit{Expansion Normalized} (EN) variables~\cite{B_hmer_2016,Bahamonde:2017ize}:
\begin{align} \label{ENvariables}
    x=\frac{\kappa \, \dot{b}}{\sqrt{6}H} \ \ \text{,} \ \  y=\frac{\kappa\,\sqrt{\abs{V}}}{\sqrt{3}H} \,,
\end{align}
where, we can see that, by its definition, the variable $y$ is always non-negative, $y\geq 0$. This provides a (by-definition) constraint on the physical phase space. Squaring \eqref{ENvariables}, we obtain: 
\begin{align}\label{e16}
   \dot{b}^2= \frac{6x^2 H^2}{\kappa^2} \ \ \text{and} \ \  |V|=\frac{3 y^2 H^2}{\kappa^2}\,.
\end{align}
We can assume that $V>0$, since we are solely interested in de-Sitter-like spacetimes. Also, we introduce the parameters of our problem\,:
\begin{align}\label{gd}
    \gamma \equiv \frac{\Lambda_0^3}{\kappa \,\Lambda_1^4} \ \ \text{and} \ \ \delta \equiv \frac{1}{\kappa\,f_b}\,,
\end{align}
where the $\gamma$ parameter gives the relative scale between the linear and cosine terms of the potential \eqref{pot}.
Another constraint on the physical phase space comes from  the Friedmann equation \eqref{friedmann1} (Friedmann\,-\,Hamiltonian constraint), which takes the following form:
\begin{align}\label{Friednmann_in_EN}
    x^2 +y^2 =1 \,.
\end{align}
With the above variables, we can finally express the equations~(\ref{friedmann1})--(\ref{KG_frw}) as a dynamical system of ordinary differential equations (ODE), as we will demonstrate below. In order to derive the equations for $x^{\prime}$ and $y^{\prime}$, where the prime denotes differentiation with respect to the quantity $N=\log\alpha (t)$, 
we use eq. \eqref{friedmann2}, and obtain\,:

\begin{align}\label{friedman_}
      & 2\dot{H} + 3H^2 = -\kappa^2\left(\frac{\dot{b}^2}{2}-V(b)\right) \Rightarrow \frac{2}{3} \,  \frac{\dot{H}}{H^2} +1 = - \frac{\kappa ^2 \,\dot{b}^2}{6 H^2} + \frac{\kappa^2 \,V }{3 H^2}   \nonumber\\
      & \Rightarrow \quad
      \frac{\dot{H}}{H^2}=-\frac{3}{2}\left(x^2 -y^2 +1\right)\,.
\end{align}
We now consider the derivative of $x$ and $y$ with respect to $N=\log\alpha$\,: 
\begin{align}
    x^{\prime}\equiv\frac{dx}{dN}=\frac{1}{H}\frac{dx}{dt}=\frac{1}{H}\left(\frac{\kappa \ddot{b}}{\sqrt{6}H} - \frac{\kappa \dot{b}\dot{H}}{\sqrt{6}H^2}\right)=\frac{\kappa \dot{b}}{\sqrt{6}H}\left(\frac{\ddot{b}}{\dot{b}H} - \frac{\dot{H}}{H^2}\right)=x \left(\frac{\ddot{b}}{\dot{b}H} - \frac{\dot{H}}{H^2}\right).
    \label{x_prime}
\end{align}
In the last equality in \eqref{x_prime}, the two terms in the brackets can be calculated by using the equation of motion for the axion \eqref{KG_frw} and also \eqref{friedman_}. From the Klein-Gordon equation, we get\,:
\begin{align*}
    \ddot{b} + 3H \dot{b}=-V_{,b} \Rightarrow \frac{\ddot{b}}{\dot{b}H}=-3 -\frac{V_{,b}} {\dot{b}H}\,.
\end{align*}
Using the definitions of the variables \eqref{ENvariables} we have:
\begin{align}\label{eom_EN}
   \frac{\ddot{b}}{\dot{b}H}  & =-3 - \frac{\kappa\,V_{,b}}{\sqrt{6}\,x\,H^2} \stackrel{\eqref{e16}}{=}  -3 - \sqrt{\frac{3}{2}} \, \frac{y^2 \,V_{,b}} {x\kappa\,V} \equiv -3 + \sqrt{\frac{3}{2}} \, \frac{y^2} {x} \, \lambda
\end{align}
where we have defined the variable\,:
$$
\lambda \equiv -\frac{V_{, b}}{\kappa V}\,.
$$
Thus, on substituting eqs.\eqref{friedman_} and \eqref{eom_EN} into \eqref{x_prime}, we obtain the equation for $x^\prime$:
\begin{align}
 x^{\prime}=-\frac{3}{2}\left[2 x-x^3+x\left(y^2-1\right)-\sqrt{\frac{2}{3}} \,\lambda \,y^2\right]\,.
\end{align}
Following similar steps presented as in \cite{Bahamonde:2017ize}, \cite{Dorlis:2024yqw} we easily arrive at the following ODE dynamical system\,:
\begin{align}\label{23}
\begin{aligned}
& x^{\prime}=-\frac{3}{2}\left[2 x-x^3+x\left(y^2-1\right)-\sqrt{\frac{2}{3}} \,\lambda \,y^2\right] \\
& y^{\prime}=-\frac{3}{2} y\left[-x^2+y^2-1+\sqrt{\frac{2}{3}} \,\lambda\, x\right] \\
& \lambda^{\prime}=-\sqrt{6}\,(\Gamma-1) \lambda^2 x
\end{aligned}
\end{align}
where we have defined:
$$
\Gamma \equiv \frac{V V_{, b b}}{V_{, b}^2}\,.
$$
The above dynamical system \eqref{23} is invariant under the transformation $y \mapsto-y$, so the dynamics for negative values of y would be a copy
of the one for positive values\,\cite{Bahamonde:2017ize}. Note also that we are assuming $H > 0$ in order to describe an expanding universe, of interest to us here. In general as $\Gamma$, $\lambda$ are functions of $b$ we can write $\Gamma= \Gamma(\lambda)$, provided that the function $\lambda(b)$ is invertible so that we can obtain $b(\lambda)$, and thus write $\Gamma(b(\lambda))$. Unfortunately, our $\lambda(b)$ is not invertible, as will be shown below, and, as a consequence, this approach fails to close the equations to an autonomous system. 
Indeed, this follows by first noting that the system \eqref{23} can be reduced, due to the Friedmann constraint \eqref{Friednmann_in_EN}, to a two-dimensional (2D) system of ODE, by eliminating the $y$ variable \ :
\begin{align}\label{24}
\begin{aligned}
& x^{\prime}=(x^2-1)\left[3x-\sqrt{\frac{3}{2}} \,\lambda \right] \\
& \lambda^{\prime}=-\sqrt{6}\,(\Gamma-1) \lambda^2 x  \,.
\end{aligned}
\end{align}
This relation holds for a general potential, within the EN formalism. In our choice of potential \eqref{pot}, $\lambda$ takes the following form\,:
\begin{align}\label{44}
    \lambda=-\frac{V_{, b}}{\kappa V}  =  \delta \,\frac{ \frac{\delta}{\gamma}\,{\rm sin}(\delta\,z)-1}{\frac{\delta}{\gamma}\,{\rm cos}(\delta\,z)+ \delta\,z}\ \ \ \text{with}\,\,\,z\equiv \kappa \,b \,,
\end{align}
which proves the non-invertible nature 
of $\lambda (b)$, as announced. 
This, in principle, urges us to restrict our analysis in a interval for $z$ and obviously avoid values of $z$ when the $\lambda$ is ill defined, i.e. when $V=0$ or equivalently $\frac{\delta}{\gamma}\,{\rm cos}(\delta\,z)+ \delta\,z=0$. 

Now, we define the variables\,:
\begin{align}\label{lambda0}
\widetilde{\lambda} \equiv-\frac{1}{\kappa b} \,\,\quad\text{and}\,\,\quad x \equiv \cos \varphi\,, 
\end{align}
with $\varphi \in[0, \pi]$ and $\widetilde{\lambda} \in \mathbb{R}$. On noting that:
\begin{align}\label{e29}
\widetilde{\lambda}^{\prime}\equiv\frac{d\widetilde{\lambda}}{dN}=\frac{1}{H}\frac{d\widetilde{\lambda}}{dt}=-\frac{1}{\kappa H} \frac{d}{dt}\left(\frac{1}{b}\right)= \frac{\dot{b}}{\kappa\,H\,b^2}= \frac{\sqrt{6}\, \cos \varphi\ }{\kappa^2\,b^2} \Rightarrow \widetilde{\lambda}^{\prime}= \sqrt{6}\, \cos \varphi\ \,\widetilde{\lambda}^2 \,,
\end{align}
we have the 2D autonomous dynamical system given by the equations\,:
\begin{align}
\begin{aligned}\label{ode1}
    \varphi^{\prime} & = \left[3 \cos \varphi-\sqrt{\frac{3}{2}}\,\left( \frac{\widetilde{\lambda} + \widetilde{\lambda}\,\frac{\delta}{\gamma} \, {\rm sin}\left(\frac{\delta}{\widetilde{\lambda}}\right)}{1- \frac{\widetilde{\lambda}}{\gamma}\, {\rm cos}\left(\frac{\delta}{\widetilde{\lambda}}\right)} \right) \,\right] \,\sin \varphi\,, \\
      \widetilde{\lambda}^{\prime}  &= \sqrt{6}\, \cos \varphi\,\,\widetilde{\lambda}^2 \,. 
      \end{aligned}
\end{align}
It is worth noting, that the transformation $\delta \rightarrow - \delta$, does not change our system \eqref{ode1}, so without loss of generality we assume that $\delta \geq 0$, and we exclude the case $\delta=0$ due to a lack of physical interpretation, given that such a case corresponds to an infinite axion coupling (which would also violate  the transplanckian-censorship conjecture~\cite{trans}, hence such a case is excluded automatically under this assumption). From now on, therefore, we impose the restriction $\delta \gtrsim 1$.

Furthermore, the system~\eqref{ode1} is not invariant under the simultaneous transformations ($\varphi \rightarrow \pi-\varphi,\, \widetilde{\lambda} \rightarrow-\widetilde{\lambda}$). So we have generally four distinctive cases $\widetilde{\lambda} > 0$ and $\gamma>0$\,, $\widetilde{\lambda} \geq 0$ and $\gamma<0$\,, $\widetilde{\lambda} \leq 0$ and $\gamma>0$\,, $\widetilde{\lambda} < 0$ and $\gamma<0$ (obviously the case when $\widetilde{\lambda}=0$ corresponds to that of $\kappa \,b \rightarrow \pm \infty$ which has been fully studied in \cite{Dorlis:2024yqw}, whilst $\gamma=0$ corresponds to the purely cosine case, see Appendix \ref{app2}). 
We note next that, if we consider the simultaneous transformations ($\varphi \rightarrow \pi-\varphi,\, \widetilde{\lambda} \rightarrow-\widetilde{\lambda}, \,\gamma \rightarrow - \gamma$), the dynamical system \eqref{ode1} is invariant. This means that we can actually just study two of the above cases, and obtain the rest by a mere reflection transformation. 
In what follows, therefore, inspired by the well studied case of \cite{Dorlis:2024yqw}, 
we first choose to analyze the models with $\widetilde{\lambda} \geq 0$ and $\gamma <0$. Then, we will study the case with $\widetilde{\lambda} < 0$ and $\gamma <0$.

\section{Dynamical System for the case \texorpdfstring{$\widetilde{\lambda} \geq 0$ and $\gamma <0$}{lambda0 > 0 and gamma < 0} and its stability analysis}\label{sec:3}
We bound the phase space through the following change of variable\,:
\begin{align}\label{zetadef}
\zeta=\frac{\widetilde{\lambda}}{\widetilde{\lambda}+1} \Rightarrow \widetilde{\lambda}=\frac{\zeta}{1-\zeta}\,,
\end{align}
which takes on values\,in\,the region $\zeta \in[0,1)$, for $\widetilde{\lambda} \in[0,+\infty)$. Thence, the system of ODE \eqref{ode1} reads\,:
\begin{align}\label{ode2}
\begin{aligned}
\varphi^{\prime} & = \left[3 \cos \varphi-\sqrt{\frac{3}{2}}\,\left( \frac{\zeta + \zeta\,\frac{\delta}{\gamma} \, {\rm sin}\left(\frac{\left(1-\zeta\right)}{\zeta} \,\delta \right)}{1-\zeta - \zeta\,\frac{1}{\gamma}\, {\rm cos}\left(\frac{\left(1-\zeta\right)}{\zeta} \,\delta \right)} \right) \,\right] \,\sin \varphi\, \\
      \zeta^{\prime}  &= \sqrt{6}\, \cos \varphi\,\,\zeta^2 \,,
\end{aligned}
\end{align}
where $\varphi \in[0, \pi]$  and \,$\zeta \in[0,1)\, ({\it cf.} \eqref{lambda0})$. 

\subsection{Stability analysis}\label{sec:3.1}
Now we want to study the stability of the system \eqref{ode2}. To this end, we should first determine the {\it critical (fixed) points} of our dynamical system, {\it i.e.} the points 
$(\varphi, \zeta)$ which satisfy\,: 
\begin{align}\label{fp}
\varphi^{\prime}= \zeta^{\prime} =0\,.
\end{align}
It is not difficult to see that, in order  
to satisfy \eqref{fp}, 
we are forced, from the $\zeta^{\prime}$ equation of \eqref{ode2},  to have either $\varphi=\pi/2$ or $\zeta=0$, or both. 

For $\zeta=0$, and $\varphi$ generic, we obtain, on substituting onto the $\varphi^{\prime}$ equation of \eqref{ode2}\,:
\begin{align}
     \cos \varphi \,\sin \varphi =0 \Rightarrow \varphi= \pi/2 \ \ \text{or} \ \ \varphi= \pi \ \ \text{or} \ \ \varphi= 0 \,,\nonumber
\end{align}
and hence we have the critical points $(\varphi,\zeta)$ to be \,${\rm O_1}=(0,0)\, ,\ {\rm A_1}=(\pi/2 ,0)\, ,\ {\rm B_1}=(\pi ,0)$\,.

For the option $\varphi=\pi/2$, and $\zeta$ generic, on the other hand, 
the $\varphi^{\prime}$ equation of \eqref{ode2} yields \ :
\begin{align}\label{43}
    \zeta + \zeta\,\frac{\delta}{\gamma} \, {\rm sin}\left(\frac{\left(1-\zeta\right)}{\zeta} \,\delta \right) =0 \,,
\end{align}
which leads to two families of solutions\,:
\begin{align}\label{twofam}
\zeta_1 =\frac{\delta}{\pi+\delta+\operatorname{arcsin}\left(\frac{\gamma}{\delta}\right)+2 \pi c_1} \ \ \ \text{and} \ \ \ \zeta_2 =\frac{\delta}{\delta-\operatorname{arcsin}\left(\frac{\gamma}{\delta}\right)+2 \pi c_2} \ \ \  \text {with }\, c_{1,2} \in \mathbb{Z}\,.
\end{align}
Note that we are obliged to assume that $\abs{\gamma/\delta }\leq 1 $, and we have further assumed that $\zeta\,,\gamma\,,\delta \neq 0$,\footnote{The case $\zeta=0, \varphi=\pi/2$ has been studied above, so for the second option we necessarily assume that $\zeta \ne 0$.} and that the denominator of $\zeta_{1,2}$ is  non-zero, in order to avoid ill-defined $\zeta_{1,2}$. Thus, we have also the infinite critical points ${\rm C_{\,c_1}}(\gamma, \delta) = \left(\pi/2,\zeta_1 \right)$ and ${\rm D_{c_2}}(\gamma, \delta) = \left(\pi/2,\zeta_2\right)$. Of course, $\zeta_{1,2}$ must lie in the interval $(0,1)$ and this imposes restrictions on $c_{1,2}$\,. 
On noting that the case $\zeta=0$, which we have examined before, corresponds to a smooth limit $c_{1,2} \to +\infty$, we can extend the range of validity of $\zeta_{1,2}$ to the interval $[0,1)$. In that case, for $\zeta_1$ we have\,:
\begin{align}
    0 \leq \zeta_1 < 1 &\Rightarrow \frac{\delta}{\pi+\delta+\operatorname{arcsin}\left(\frac{\gamma}{\delta}\right)+2 \pi c_1} < 1  \Rightarrow \pi+\operatorname{arcsin}\left(\frac{\gamma}{\delta}\right)+2 \pi c_1> 0 \nonumber \\ &\stackrel{(\gamma=-\delta)}{\Rightarrow} c_1 > - \frac{1}{4}  \Rightarrow c_1 \in  \mathbb{N}\cup\{0\}\,,
\end{align}
while for $\zeta_2$ we obtain\,:
\begin{align}
    0 \leq \zeta_2 < 1 &\Rightarrow \frac{\delta}{\delta-\operatorname{arcsin}\left(\frac{\gamma}{\delta}\right)+2 \pi c_2} < 1 \Rightarrow  2 \pi c_2 > \operatorname{arcsin}\left(\frac{\gamma}{\delta}\right) \nonumber \\ &\stackrel{(\gamma=-\delta)}{\Rightarrow} c_2 > -\frac{1}{4} \Rightarrow c_2 \in \mathbb{N}\cup\{0\}\,.
\end{align}

\begin{figure}[ht!]
    \centering
\includegraphics[width=\textwidth]{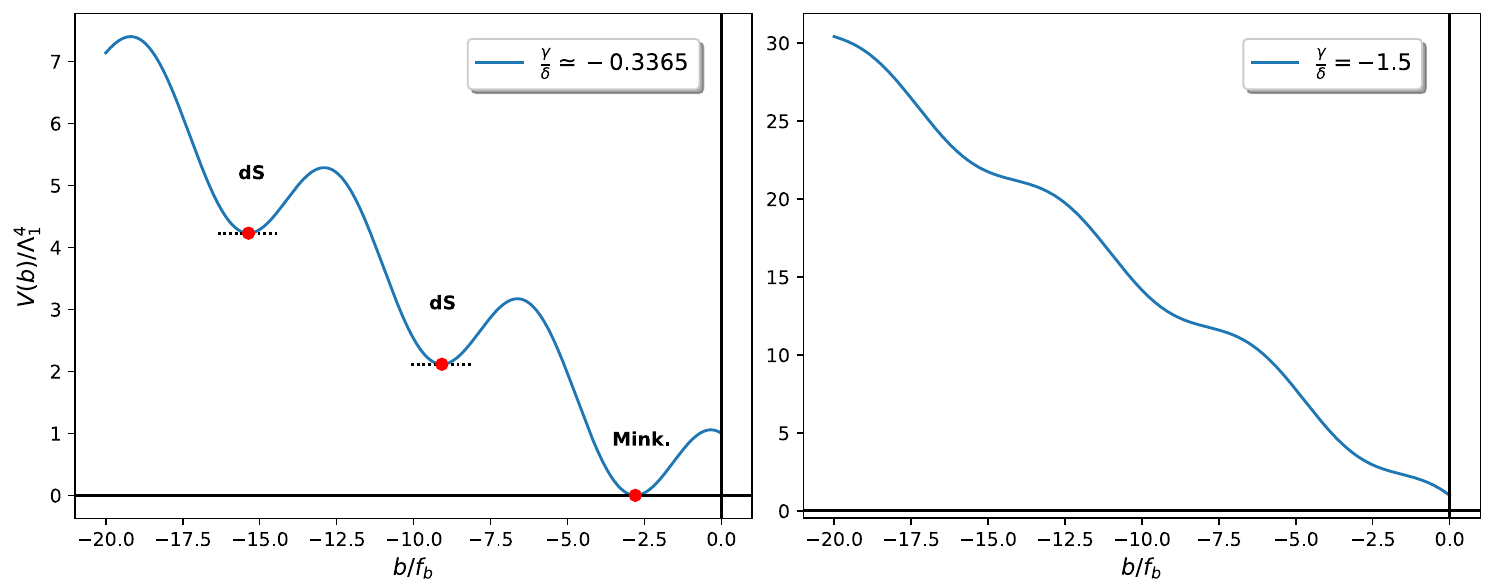}
    \caption{\underline{Left panel}: Only de-Sitter vacua appear for $-1 \le \gamma/\delta < -0.3365$ and $b/f_b<0$. The series of vacua correspond to a decreasing effective cosmological constant, with a Minkowski end-point for $\gamma/\delta\simeq-0.3365$. \underline{Right panel}: For  $\abs{\gamma/ \delta} \geq 1$, the linear term dominates and  periodic modulations appear due to the instanton induced cosine potential. }
    \label{figure0}
\end{figure}

It is instructive to see how the aforementioned two families intuitively correspond to the critical points of the original potential. Essentially, as we want de-Sitter-like spacetimes, we must impose the condition $V>0$ for all values of $b/f_b$. For our potential \eqref{pot} we have that\,:
\begin{align}
     & V_{,b}=0 \Rightarrow  {\rm sin}\left(\frac{b}{f_b}\right)=\frac{\Lambda_{0}^3\,f_b }{\Lambda_1^4} \equiv \frac{\gamma}{\delta} \nonumber \\ &\Rightarrow \frac{b}{f_b}=\pi-\operatorname{arcsin}\left( \frac{\gamma}{\delta} \right)-2 \pi c_1 \,\,\text{or}\,\,\frac{b}{f_b}=\operatorname{arcsin}\left( \frac{\gamma}{\delta} \right)-2 \pi c_2 \nonumber\,,
\end{align}
where $c_{1,2} \in  \mathbb{N}\cup\{0\}$, and we call these two solutions $b_{1,2}$, respectively. Thus, there are two families of solutions of critical points, in accordance with the two families of critical points of the dynamical system \eqref{ode2}, by means that there is an one to one mapping between them. The first family is a family of local minima because\,:
\begin{align}
     \left.V_{,bb}\right|_{b=b_1} &= - \frac{ \Lambda_1^4}{f_b ^2} \, {\rm cos}\left(\pi-\operatorname{arcsin}\left( \frac{\gamma}{\delta} \right)-2 \pi c_1\right) 
     \nonumber \\ &=\frac{ \Lambda_1^4}{f_b ^2} \, {\rm cos}\left(\operatorname{arcsin}\left( \frac{\gamma}{\delta} \right)\right) = \frac{ \Lambda_1^4}{f_b ^2} \sqrt{1-\left( \frac{\gamma}{\delta} \right)^2} \geq 0 \nonumber\,.
\end{align}
Hence, the first family corresponds to minima. Similarly, for the second family of fixed points, we have\,:
\begin{align}\label{vbb2}
     \left.V_{,bb}\right|_{b=b_2} &= - \frac{ \Lambda_1^4}{f_b ^2} \, {\rm cos}\left(\operatorname{arcsin}\left( \frac{\gamma}{\delta} \right)-2 \pi c_1\right) \nonumber \\&=-\frac{ \Lambda_1^4}{f_b ^2} \, {\rm cos}\left(\operatorname{arcsin}\left( \frac{\gamma}{\delta} \right)\right) = -\frac{ \Lambda_1^4}{f_b ^2} \sqrt{1-\left(\frac{\gamma}{\delta} \right)^2} \leq 0  \,.
\end{align}
Therefore, the second family corresponds to maxima\footnote{\label{foot3}The equality in \eqref{vbb2} corresponds to the limit $\abs{\gamma/\delta}=1$, in which the two families ${\rm C_{\,c_1}}(\gamma, \delta)$\,,\,${\rm D_{\,c_2}}(\gamma, \delta)$ degenerate into one, with the points being strictly decreasing points of inflection, as can be inferred by computing the third derivative of the potential \eqref{pot}. In the dynamical-system formalism, these points are saddle fixed points (see subsection \ref{subsec1AppC} of Appendix \ref{app3} for details).}. We restrict ourselves to the case 
$V>0$, so that the analysis is valid (implying only de Sitter-type vacua ({\it cf.} figure \ref{figure0})). 
This leads to  $V(b)/\Lambda_1^4 >0$ at the minimum where $c_1=0$ and corresponds to the first family (\,namely ${\rm C_{\,c_1}}(\gamma, \delta)= \left(\pi/2,\,\zeta_1 \right)$\,). We impose\,:
\begin{align}
\begin{aligned}\label{49}
     \left.\frac{V(b)}{\Lambda_1^4}\right|_{b=b_1} &= \frac{\gamma}{\delta}\,\left( -\pi-\operatorname{arcsin}\left( \frac{\gamma}{\delta} \right) \right) + {\rm cos}\left( -\pi-\operatorname{arcsin}\left( \frac{\gamma}{\delta} \right) \right)  \\&=  \frac{\gamma}{\delta}\,\left( -\pi-\operatorname{arcsin}\left( \frac{\gamma}{\delta} \right) \right) - \sqrt{1-\left( \frac{\gamma}{\delta} \right)^2} >0 \,.
     \end{aligned}
\end{align}

A study of this function of $\gamma/\delta$, reveals that it is positive only when  
\begin{align}\label{lowerbound}
  \frac{\gamma}{\delta}\,\in [-1\,,\,-0.3365)\,,
\end{align}
for which the critical points correspond to de Sitter expansion.\footnote{Of course, it should be remarked that the root of equation \eqref{49} is a transcendental number and as such, it can be determined in as many digits as the desired accuracy requires.}

A possible physical interpretation of the series of de Sitter vacua depicted in figure~\ref{figure0} (left panel) is given at the end of section~\ref{sec:5}. Also in figure \ref{figure0} (right panel), we depict the potential in the case when $\abs{\gamma/ \delta} \geq 1$. This case is to be understood as the case in which the potential becomes approximately linear with minor modulations.

The Jacobian (stability) matrix needs to be calculated so as to study the stability of the critical points. At each hyperbolic fixed point the stability can be determined from the eigenvalues of the Jacobian directly, and those eigenvalues have non-zero real part. If one eigenvalue is zero, corresponding to a center space, more advanced techniques such as center manifold theory should be implemented in order to draw safe conclusions on the stability of the critical points and these points are called non-hyperbolic critical points with one eigenvalue having zero real part~\cite{Bahamonde:2017ize,boehmer2010,Boehmer:2011tp}.
The Jacobian (stability) matrix reads\,:
\begin{align}
J( \varphi, \zeta) =
\begin{bmatrix}
    \dfrac{\partial \varphi^{\prime}}{\partial \varphi} & \dfrac{\partial \varphi^{\prime}}{\partial \zeta} \\[1.5ex]
     \dfrac{\partial \zeta^{\prime}}{\partial \varphi} & \dfrac{\partial \zeta^{\prime}}{\partial \zeta}
\end{bmatrix}
= \begin{bmatrix}
 \mathcal A(\gamma, \delta, \zeta, \phi) -3 \sin^2\varphi+3 \cos ^2 \varphi  & \quad {\mathcal B}(\gamma, \delta, \zeta, \phi)  \\
 -\sqrt{6} \zeta ^2 \sin \varphi  & \quad 2 \sqrt{6} \zeta  \cos \varphi  \\
 \end{bmatrix}
\end{align}
where $$
 A(\gamma, \delta, \zeta, \varphi) \equiv \frac{\sqrt{6} \, \zeta \, \cos \varphi \, \left(\gamma +\delta  \sin \left(\frac{\left(1-\zeta\right)}{\zeta} \,\delta \right)\right)}{2 \gamma  (\zeta -1)+2 \zeta  \cos \left(\frac{\left(1-\zeta\right)}{\zeta} \,\delta \right)}\,,$$ 
 and 
 $${\mathcal B}(\gamma, \delta, \zeta, \varphi) \equiv -\frac{\sqrt{\frac{3}{2}} \sin\varphi \left(\zeta  \left(\gamma ^2+2 \gamma  \delta  \sin \left(\frac{\left(1-\zeta\right)}{\zeta} \,\delta \right)+\delta ^2\right)+\gamma  \delta ^2 (\zeta -1) \cos \left(\frac{\left(1-\zeta\right)}{\zeta} \,\delta \right)\right)}{\zeta  \left(\gamma  (\zeta -1)+\zeta  \cos \left(\frac{\left(1-\zeta\right)}{\zeta} \,\delta \right)\right)^2}\,.$$
Now, we want to find the eigenvalues of the Jacobian matrix at the critical points. The points ${\rm O_1}$, ${\rm A_1}$, ${\rm B_1}$ corresponding to $\zeta=0$, $\widetilde{\lambda} =0$, are fully described by the dynamical system with a linear axion potential analyzed in \cite{Dorlis:2024yqw}, given that the limit $\zeta \rightarrow 0$ is equivalent to $\abs{\gamma} \rightarrow \infty$. Thus, these fixed points are unstable, saddle or unstable, respectively. For a study of the two families of critical points, viz. ${\rm C_{\,c_1}}(\gamma, \delta)$,\,${\rm D_{c_2}}(\gamma, \delta)$ the straightforward approach is somewhat tedious, and, as such, some technical aspects are presented in Appendix \ref{appB}, where we provide the stability analysis. Essentially, the ${\rm C_{\,c_1}}(\gamma, \delta)$ family corresponds to stable spiral\,/\,node fixed points whereas the ${\rm D_{c_2}}(\gamma, \delta)$ to saddles. For example, by choosing $\gamma=-3.7$ and $\delta=10$, the stability of the two families, for different values of $c_1\,,c_2 \in \mathbb{N}\cup\{0\}$, is depicted in Tables \ref{1stfamily}, \ref{2ndfamily}, where the above claims become clear. The behavior of the dynamical system \eqref{ode2}, as far as the critical points and their stability are concerned, is summarized in Table \ref{table_sum}.

\begin{table}[ht!]
\begin{align}
\begin{tabular}{|c|c|c|c|c|}
\hline Critical Point  & $\varphi$ & $\zeta$& Eigenvalues & Stability \\
\hline${\rm C_{0}}$ & $\pi/2$ & $0.78354 $  & $-1.5-54.6867 i$\,,$-1.5+54.6867 i$& stable spiral\\
\hline${\rm C_{1}}$ & $\pi/2$ & $0.525051 $  &\, $-1.5-10.631 i$\,,$-1.5+10.631 i$ & stable spiral\\
\hline$\vdots$ & $\vdots$ &$\vdots$ &\, $\vdots$\,& $\vdots$\\
\hline${\rm C_{53}}$ & $\pi/2$ & $0.0289208 $  & $-1.5-0.101493 i$\,,$-1.5+0.101493 i$& stable spiral\\
\hline${\rm C_{54}}$ & $\pi/2$ & $0.0284047 $  & $-1.67755$\,,$-1.32245$ & stable node\\
\hline$\vdots$ & $\vdots$ &$\vdots$ &\, $\vdots$\,& $\vdots$\\
\hline${\rm C_{10000}}$ & $\pi/2$ & $0.000159123 $  & $-2.996$\,,$-0.00400194$ & stable node\\
\hline$\vdots$ & $\vdots$ &$\vdots$ &\, $\vdots$\,& $\vdots$\\
\hline
\end{tabular}\nonumber
\end{align}
\caption{Stability for first family ${\rm C_{\,c_1}}(\gamma, \delta)$ with $\gamma=-3.7$ and $\delta=10$.}
\label{1stfamily}
\end{table}

\begin{table}[ht!]
\begin{align}
\begin{tabular}{|c|c|c|c|c|}
\hline Critical Point  & $\varphi$ & $\zeta$& Eigenvalues & Stability \\
\hline${ \rm D_{0}}$ & $\pi/2$ & $0.963483 $  &\, $-17.7144$\,,$14.7144$\,& saddle\\
\hline${\rm D_{1}}$ & $\pi/2$ & $0.600161 $  &\, $-10.6852$\,,$7.68516$\,& saddle\\
\hline$\vdots$ & $\vdots$ &$\vdots$ &\, $\vdots$\,& $\vdots$\\
\hline${\rm D_{10000}}$ & $\pi/2$ & $0.000159129$  &\, $-3.00399$\,,$0.00399072$\,& saddle\\
\hline$\vdots$ & $\vdots$ &$\vdots$ &\, $\vdots$\,& $\vdots$\\
\hline
\end{tabular}\nonumber
\end{align}
\caption{Stability for second family ${\rm D_{c_2}}(\gamma, \delta)$ with $\gamma=-3.7$ and $\delta=10$.}
\label{2ndfamily}
\end{table}

\begin{table}[ht!]
\begin{center}
\begin{tabular}{|c|c|c|c|c|}
\hline \textbf{Critical Point}  & $\varphi$ & $\zeta$ & \textbf{Stability} \\
\hline${\rm O_1}$ & $0$ & $0$ & unstable   \\
\hline${\rm A_1}$  & $\pi/2$ & $0$ & saddle \\
\hline${ \rm B_1}$  & $\pi$ & $0$ & unstable \\
\hline${\rm C_{\,c_1}}(\gamma, \delta)$ & $\pi/2$ & $\zeta_1 $ for $\abs{\gamma/\delta }\leq 1 $ & stable spiral\,/\,node \\
\hline${\rm D_{c_2}}(\gamma, \delta)$ & $\pi/2$ & $\zeta_2 $ for $\abs{\gamma/\delta }\leq 1 $ & saddle \\
\hline
\end{tabular}\nonumber 
\caption{Summary of the behavior (critical points and their stability) of the dynamical system \eqref{ode2}.}
\label{table_sum}
\end{center}
\end{table}

\section{Classification of inflationary vacua for different values of \texorpdfstring{$\gamma/ \delta$}{gamma/delta}}\label{sec:5}

From the above critical-point analysis, 
it follows that our dynamical system \eqref{ode2} is characterized by qualitatively different  phase spaces, and consequently stability behavior, 
depending on whether $\abs{\gamma/\delta }\leq 1 $ or $\abs{\gamma/\delta }\geq 1 $. The former case contains the two families of critical points ${\rm C_{\,c_1}}(\gamma, \delta)$,\,${\rm D_{c_2}}(\gamma, \delta)$, while the latter, in which such families are absent, 
 is qualitatively similar to the case studied in \cite{Dorlis:2024yqw}. In other words,
at $\abs{\gamma/\delta}=1$, bifurcation of the critical points occurs~\cite{strogatz:2000} (see also footnote \ref{foot3}, as well as the detailed discussion on the nature of the stability of this case in \ref{subsec1AppC} of Appendix \ref{app3}, where it is also shown that the fixed points correspond to a family of saddle points). With these remarks in mind, we now proceed to study inflation 
 in these regimes for the potential \eqref{pot}, giving emphasis on determining phenomenologically interesting (metastable) inflationary scenarios. 
 
\subsection{Case where \texorpdfstring{$\abs{\gamma/\delta} \geq 1$}{gamma/delta}}
\begin{figure}[htbp!]
    \centering
\includegraphics[width=0.563\textwidth]{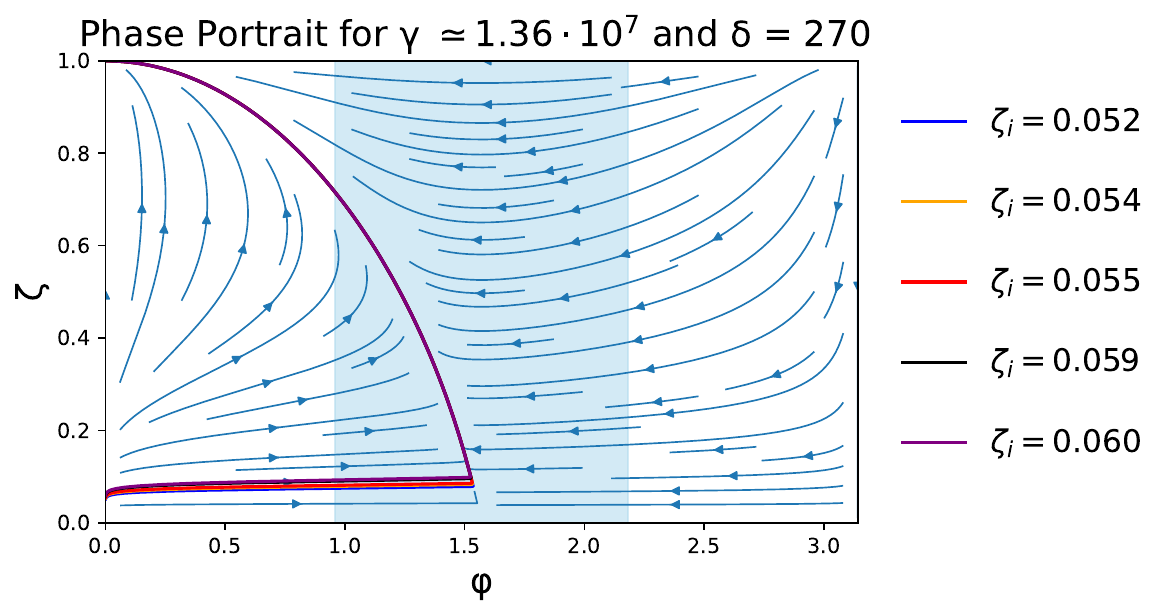}\hfil\includegraphics[width=0.437\textwidth]{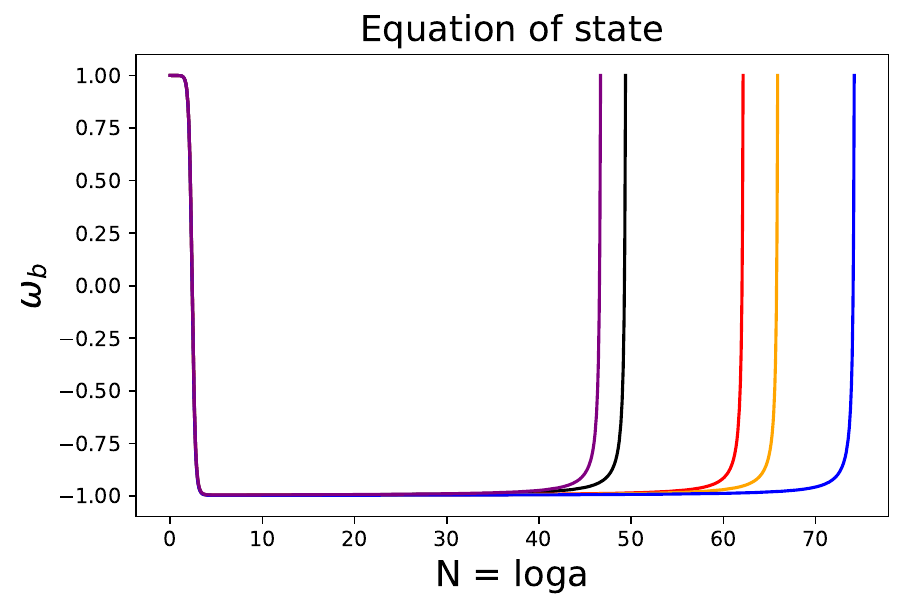}
    \caption{Phase space (left panel) and EoS (right panel) for $\varphi_i =0.001$ with different values of $\zeta_i$ and  $\abs{\gamma/\delta} \gg 1$.}
    \label{figure1}
\end{figure}

In this case only the critical points identified at infinity $\kappa \,b  \rightarrow - \infty$ exist. This case is well understood from the work in \cite{Dorlis:2024yqw}. Indeed, when $|\gamma|$ is sufficiently large compared to $\delta$, the system reduces approximately to that in\,\cite{Dorlis:2024yqw}, but with characteristic modulations in the trajectories and EoS. Since we wish to confirm the phenomenologically-relevant choices made in \cite{Dorlis:2024uei} via our dynamical system analysis, we set\,:
\begin{align}\label{48}
 \delta=270\,,\,\,\,\Lambda_{0}^3 \simeq - 0.768 \cdot10^{-7} M_{\rm Pl}^3\,,\,\,\,\Lambda_{1}^4 \simeq 5.63 \cdot 10^{-15} M_{\rm Pl}^4\,\,\Rightarrow\, \abs{\gamma}\simeq 1.36\cdot 10^7 \gg 1\,,
\end{align}
which guarantees the dominance of the linear term in the axion potential. The results are depicted 
in figure~\ref{figure1}, for different initial conditions. The shaded rectangular region in the phase space (for this and all subsequent phase portraits in this article) corresponds to an accelerated expansion, with EoS\, $\omega_b\leq-1/3$.

Since not all initial conditions can lead to an inflationary state, it is evident that a selection of the initial conditions is in order. By selecting appropriately the initial values for $\varphi$, $\zeta$, we obtain a characteristic cosmological scenario that encapsulates phenomenologically relevant results. As becomes evident from figure \ref{figure1}, for a fixed initial choice of $\varphi$, only a specific range of initial $\zeta$ give us the desired inflationary phenomenology~\cite{Planck}, {\it i.e.} a number of e-foldings in the range  ${\mathcal N}_{\rm e} = \mathcal O(50-60)$. It is worth mentioning that in our case, where modulations are present, inflation is slightly prolonged compared to 
 the linear-axion-potential~\cite{Dorlis:2024yqw}, that is, we have a larger number of e-folds with the same initial conditions\footnote{For the plots we used python and particularly, for the streamplot the matplotlib package, while  for the trajectories we implemented the Radau integration method.}. 
 
\begin{figure}[htbp!]
    \centering
\includegraphics[width=0.563\textwidth]{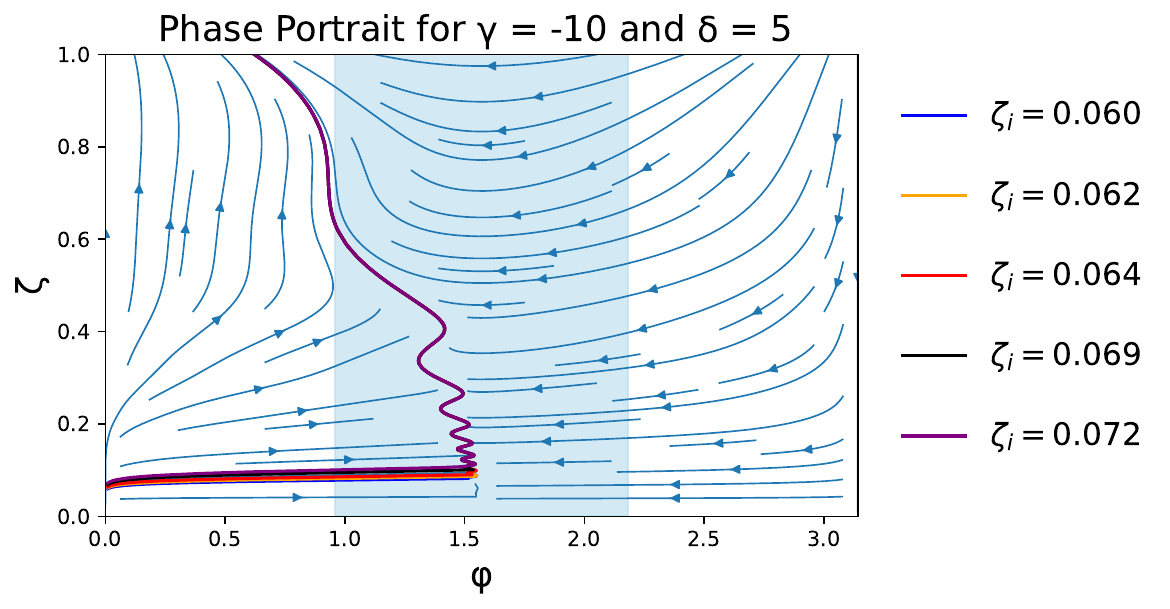}\hfil\includegraphics[width=0.437\textwidth]{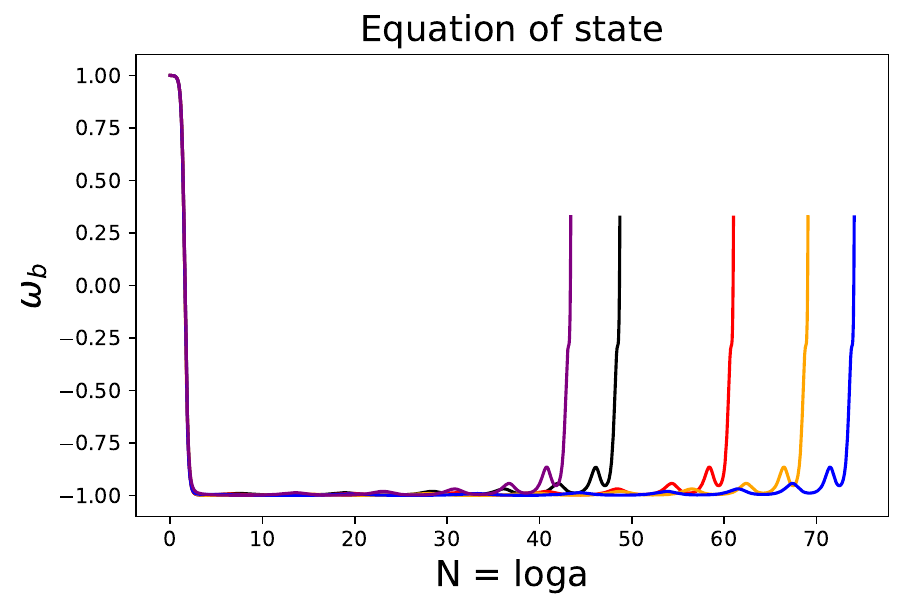}
    \caption{Phase space (left panel) and EoS (right panel) for the model corresponding to $\gamma=-10$ and $\delta=5$,
    with initial conditions $\varphi_i =0.01$ with a varying $\zeta_i$.}
    \label{figure4}
\end{figure}

 In \cite{Dorlis:2024yqw}, it was argued that the linear axion potential  emerged from the condensation of (primordial) gravitational waves. Inflation in that model was preceded by a stiff era, and thus the system was initially resting at ${\rm O_1}(0,0)$ critical point, where, basically, any small perturbation would have triggered  inflation\footnote{In the case of \cite{Dorlis:2024uei,Dorlis:2024yqw} such a small perturbation is attributed to the formation of the condensate itself which slightly alters the equation of state from the stiff one.}. Suitable initial conditions were required in order to obtain the correct phenomenology and a de-Sitter-like universe, which remarkably lead to similar results as in 
 the StRVM approach of \cite{bms,ms1,ms2}. In the modulated case studied in this article, initial conditions close to the $\zeta=0$ line also give us inflation. In the aforementioned regime of parameters \eqref{48}, the modulations in the phase space and EoS exist, but are extremely small, since the linear term in the axion potential \eqref{pot} dominates  the cosine one by many orders of magnitude. So, it is instructive to see the profiles of other, not so extreme, cases of $\abs{\gamma/\delta} \ge 1$. For instance, we pick $\delta=5$ and $\gamma=-10$ ({\it cf.} figure \ref{figure4}), as a characteristic case. Now the modulations are profound. 
In this case, for the initial condition $\varphi_i =0.01$, the correct phenomenology of ${\mathcal N}_{\rm e} = \mathcal O(50-60)$ is obtained only for the range $\zeta_i = [0.064\,,\,0.069]$. Moreover, we cannot have $\abs{\gamma}<1$, since we must preserve the condition $\abs{\gamma/\delta} \geq 1$ (recall that $\delta \gtrsim 1$.) Hence, the phase space cannot exhibit a purely-cosine behavior as in the $\abs{\gamma} \ll 1$ case, studied in Appendix \ref{app2}.

\subsection{Case where \texorpdfstring{$\abs{\gamma/\delta} \leq 1$}{gamma/delta}}\label{sec4.2}
\begin{figure}[ht!]
    \centering
    \includegraphics[width=\textwidth]{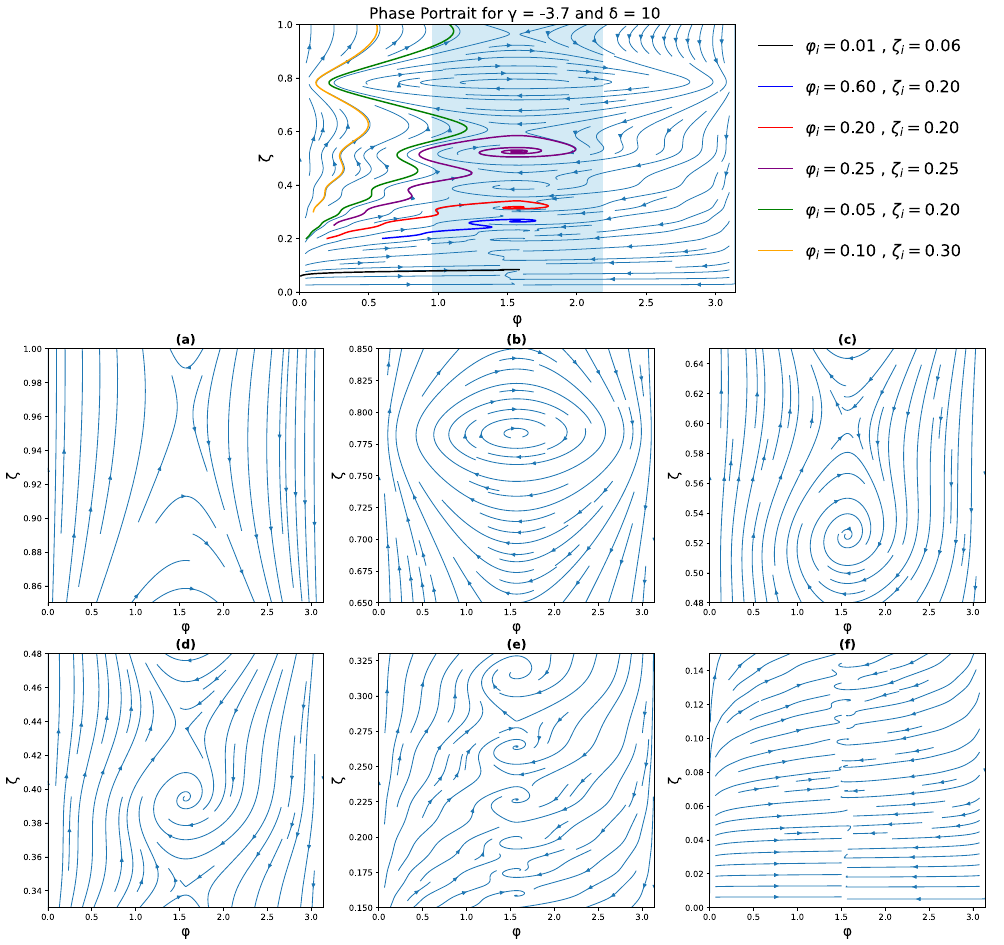}
    \caption{\underline{Upper panel}: Phase space for $\gamma=-3.7$ and $\delta=10$. There are trajectories which lead to eternal inflation classically ({\it e.g.} the purple-colored curve), while others (the green-colored curve) avoid it. \underline{Middle and Lower panels}: specific regions of the phase portrait, demonstrating the different behavior of the system at different critical points (middle panel from left to right: (a) saddle, (b) stable spiral, and (c) saddle, stable spiral fixed points, respectively; lower panel from left to right: (d) saddle and stable spiral,  (e) stable spiral and saddle, and (f) saddle, stable-node fixed points, respectively).}
    \label{fig:7}
\end{figure}

\begin{figure}[ht!]
    \centering
    \includegraphics[width=0.8\textwidth]{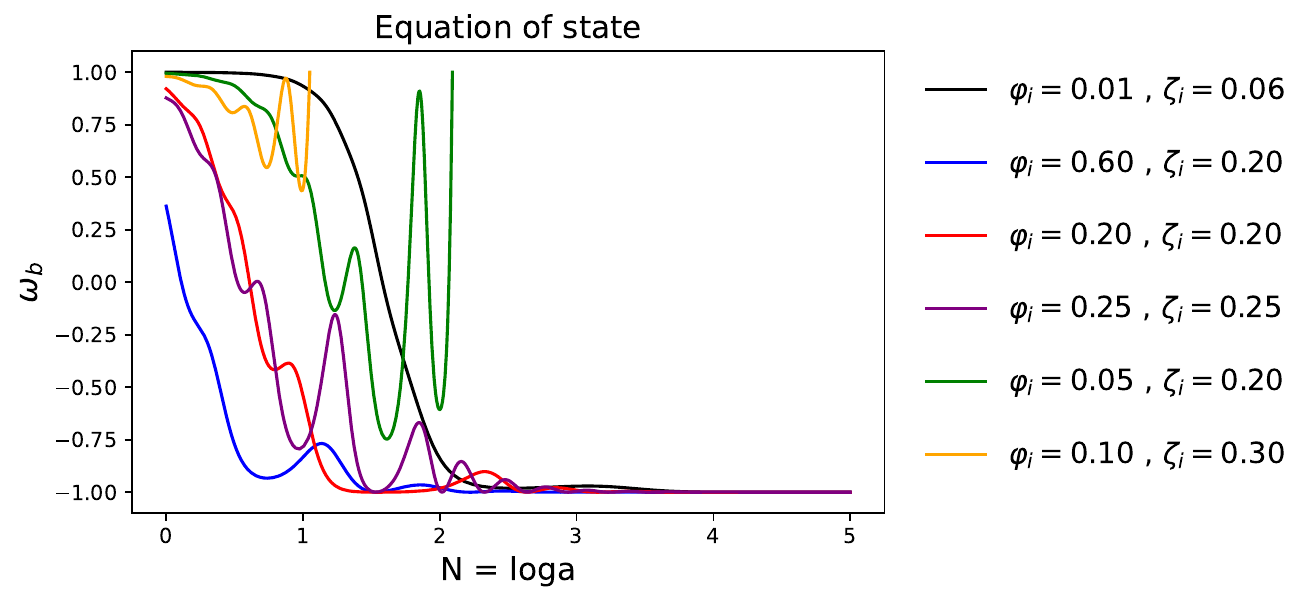}
    \caption{EoS for different values of $\varphi_i$\,, $\zeta_i$ for $\gamma=-3.7$ and $\delta=10$.}
    \label{fig:9}
\end{figure}

In this case, besides the critical points at infinity, we also have an infinite number of fixed points that belong to the two families ${\rm C_{\,c_1}}(\gamma, \delta)$ and ${\rm D_{c_2}}(\gamma, \delta)$. The phase space displays different qualitative behavior compared to the case studied in the previous subsection. We choose for concreteness the case $\delta=10$ and $\gamma=-3.7$. We depict the situation for the relevant phase spaces in figures \ref{fig:7} and \ref{fig:9}, focusing our attention on specific regions of the phase portraits that reveal the distinctive nature of the different critical points in the aforementioned families.

In the phase space we have saddle and stable (spiral\,/\,node) fixed points. 
The latter category implies that, under appropriate initial conditions, one can obtain cosmological settings with eternal inflation, provided that the trajectory is attracted to such a stable critical point. On the other hand, as becomes clear from figure~\ref{fig:7}, under different initial conditions the system transits to another phase, without passing through inflation.

\begin{figure}[htbp!]
    \centering
\includegraphics[width=0.49\textwidth]{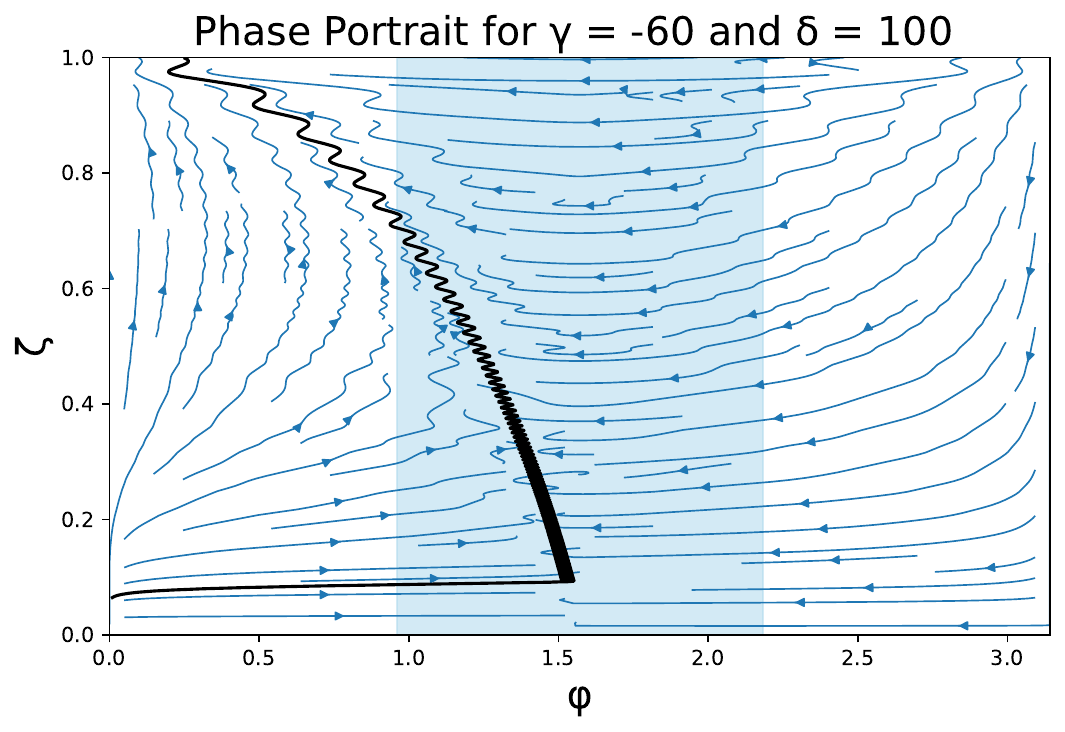}\hfil\includegraphics[width=0.51\textwidth]{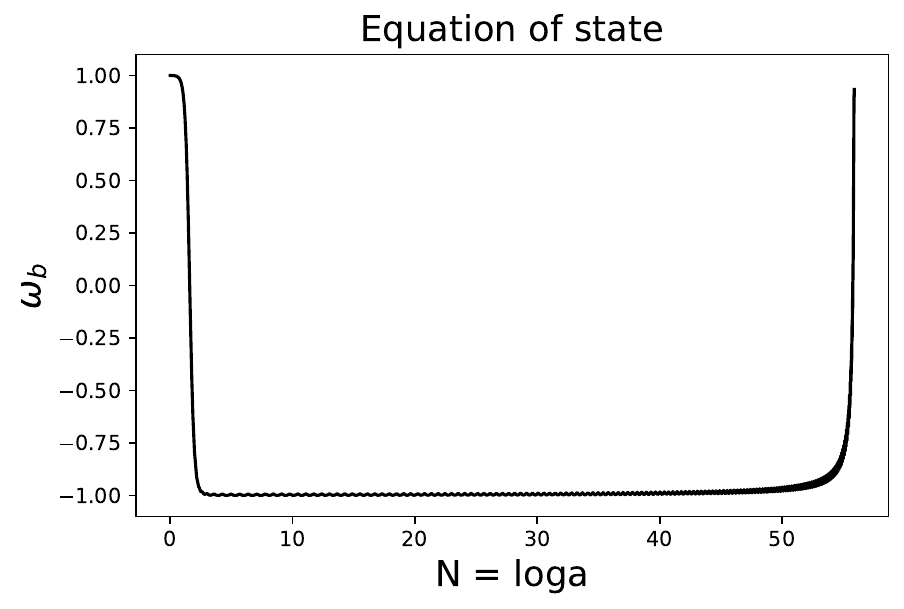}
    \caption{Phase space (left panel) and EoS (right panel) for initial conditions $(\varphi_i =0.01, \zeta_i =0.065)$, in the model with $\gamma=-60$ and $\delta=100$.}
    \label{fig:10}
\end{figure}

In addition, we can find appropriate $\gamma\,,\delta$ with $\abs{\gamma/\delta} \leq 1$ such that inflation is realized, with relevant initial conditions. This means that we can find initial conditions for which inflation is achieved and is characterized by a graceful exit, with the phenomenologically-relevant duration~\cite{Planck} ${\mathcal N}_{\rm e} = \mathcal O(50-60)$, for instance by choosing $\gamma=-60$ and $\delta=100$. The relevant phase portraits are depicted in figure \ref{fig:10}.

This should come as no surprise, since in this case despite $\abs{\gamma/\delta} \leq 1$, with the value  $\abs{\gamma} =60 \gg 1$, one obtains the dominance of the linear axion term in the potential \eqref{pot}, but with significant periodic modulations.

A remark is in order at this point concerning the limiting case $\abs{\gamma/\delta} \rightarrow 1-\epsilon$ with $\epsilon \rightarrow 0^{+}$. In this limit the pertinent phase portrait changes qualitatively, and in this sense we have a bifurcation curve in the phase space (see figures \ref{figure1},\,\ref{fig:7} upper panel). 

We also remark, for completeness, that in the case $\abs{\gamma/\delta} \leq 1$, there is no GW-induced condensate inflation of the type discussed in \cite{Dorlis:2024yqw}. This is due to the fact that the pertinent equations of motion of the $b$-axion field imply a constant axion, $
\dot b =0$, and hence, a vanishing gravitational anomaly condensate (which is proportional to $\dot b$).

\begin{figure}[ht!]
    \centering
    \includegraphics[width=\textwidth]{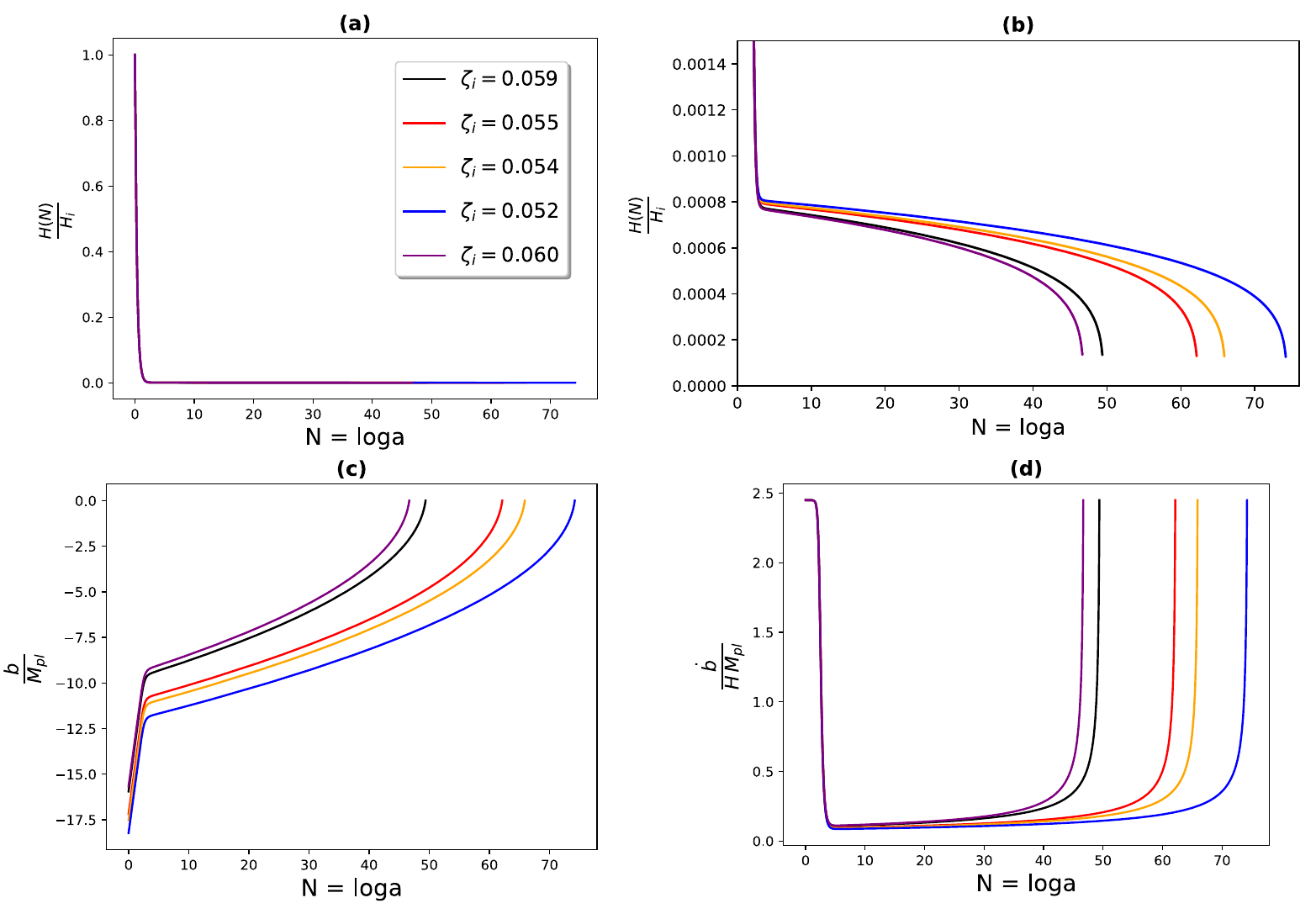}
    \caption{The plots correspond to the case in figure \ref{figure1}, with initial conditions  $\varphi_i =0.001$, and various values of $\zeta_i$. In the Upper panels (a) and (b) we illustrate the evolution of the relative change of the Hubble rate $H$ with respect to its initial value, demonstrating the approximate constancy of $H$ during inflation, which last for about ${\mathcal O}(50-60)$ e-foldings, depending on the values of $\zeta_i$. In the lower panels (c) and (d) we give the behavior of $b$ and $\dot b$, respectively, demonstrating explicitly the satisfaction of the slow-roll condition \eqref{Hdot}, on account of \eqref{friedmann2}.}
    \label{fig:new7}
\end{figure}

\begin{figure}[ht!]
    \centering
    \includegraphics[width=\textwidth]{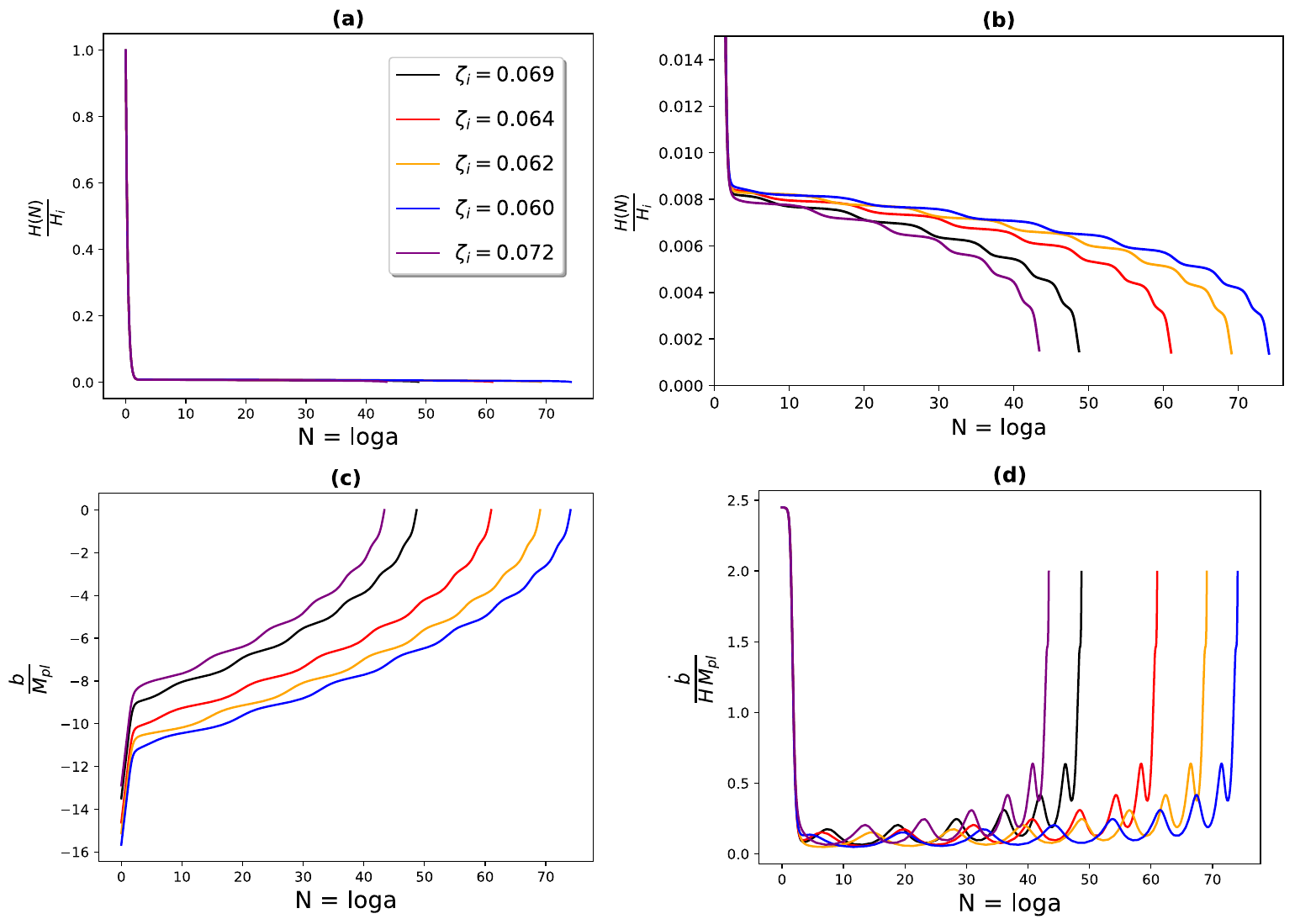}
    \caption{As in figure \ref{fig:new7}, but for the the case depicted in figure \ref{figure4}, with initial conditions $\varphi_i =0.01$, and different values of $\zeta_i$. The slow-roll conditions of inflation are demonstrated.}
    \label{fig:new8}
\end{figure}

\begin{figure}[ht!]
    \centering
    \includegraphics[width=\textwidth]{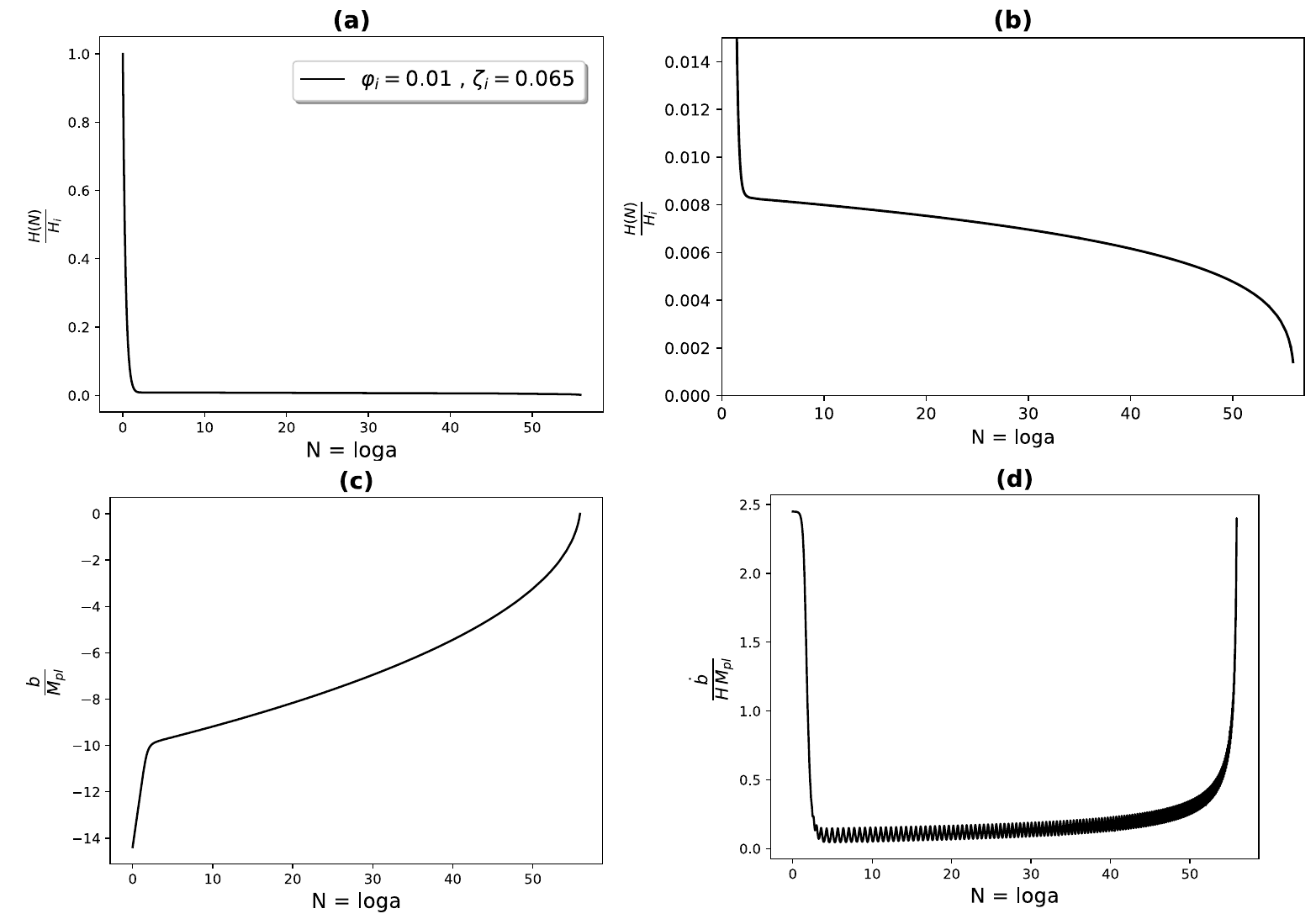}
    \caption{As in figures \ref{fig:new7}, \ref{fig:new8} but for the model depicted in 
    figure \ref{fig:10}, corresponding to the initial conditions $\varphi_i = 0.01$, $\zeta_i = 0.065$. Again, the satisfaction of the slow-roll conditions during inflation is evident.}
    \label{fig:new9}
\end{figure}

At this point we make some comments regarding the slow-roll features of the inflationary state for the aforementioned scenarios.
From \eqref{friedmann2}, one observes that a sufficient condition for the slow-roll condition \eqref{Hdot} to be satisfied is to have\, $\dot b \lesssim \mathcal O(0.1) H M_{\rm Pl}$, which was the situation encountered in \cite{Dorlis:2024yqw}. This characterizes the cases discussed above. Indeed, on using Eq.~\eqref{ENvariables}, with $y=\sin \varphi$, we observe that the evolution of the Hubble rate is given by\,:
\begin{align}\label{Hratio}
    \frac{H(N)}{H_i}= \sqrt{\frac{\gamma \,\left(1-1/\zeta     \right) + \cos \left(\delta\,\left(1-1/\zeta     \right) \right)}{\gamma \,\left(1-1/\zeta_i     \right) + \cos \left(\delta\,\left(1-1/\zeta_i     \right) \right)}}\,\frac{\sin\varphi_i}{\sin\varphi}\,,
\end{align}
where the subscript {\it i} denotes initial values. Similarly, from the definition \eqref{ENvariables} of the EN variable, and eqs.~\eqref{lambda0},\eqref{zetadef}, we find that the evolution of the $b$ field and its time derivative $\dot{b}$ read\,:
\begin{equation}\label{dbb}
    \frac{b}{M_{\rm Pl}}=1-\frac{1}{\zeta} \ , \ \ 
     \frac{\dot{b}}{H M_{\rm Pl}}= \sqrt{6} \cos\phi \,.
 \end{equation} 
The corresponding results are depicted in figures \ref{fig:new7}, \ref{fig:new8}, \ref{fig:new9}, from which we easily observe that the conditions for a slow-roll inflation are comfortably satisfied. Indeed, from figures \ref{fig:new7}(a),(b) (upper panels), we observe that the Hubble rate decreases quickly, until it assumes an approximately constant value $H_I \simeq {\rm const.}$, which characterizes a de-Sitter phase. In figures \ref{fig:new7}(c),(d) (lower panels), on the other hand, we depict the behavior of $b$ and $\dot{b}$ during inflation (which lasts ${\mathcal O}(50-60)$ e-foldings). We deduce that, since $b\,,\dot{b}$ remain in the same order of magnitude, we have the relations\,:
\begin{equation}
    \frac{\abs{b}}{M_{\rm Pl}} \sim \mathcal{O}\left(10\right) , \ \  \frac{\dot{b}}{H_I M_{\rm Pl}} \sim \mathcal{O}\left(10^{-1}\right)\,,
    \label{b_bdot_orders}
\end{equation}
similarly to the situation in~\cite{Dorlis:2024yqw}.

We now come to a discussion on the potential physical significance of the discrete series of the stable critical points of the first family 
${\rm C_{\,c_1}} (\gamma, \delta)$ depicted in figure~\ref{figure0}, each of which corresponds classically to a stable (eternal) de Sitter vacuum. 
Inspecting the form of the series of vacua, with descending values of the potential at the local minima, as the value of the field $b$ approaches zero from the left, we may conjecture that such a series describes a situation of a universe, which passes ({\it e.g.} via {\it quantum tunneling} effects) through a discrete series of de-Sitter vacua until it reaches asymptotically the vacuum that corresponds to a zero value of the potential, which could be viewed as a Minkowski equilibrium spacetime, when $\abs{\gamma/\delta}=0.3365$ \,({\it cf.} fig.\ref{figure0}). When $\abs{\gamma/\delta}>0.3365$, a remnant of a quite small effective cosmological constant can be obtained, which might be relevant for the current de Sitter phase of the universe. 

It is interesting to notice that such a discrete series of inflationary vacua has been argued also to characterize Liouville-string cosmologies~\cite{Antoniadis:1988aa}, based on unitary minimal conformal models as their ``internal'' spaces. The latter have central charges $c_n$, which approach from below accumulation points with integer or half-integer values, $c_n= c_\infty - \mathcal O(\frac{1}{n})$, with $c_{\infty} \in \frac{\mathbb{Z}}{2}$. 
In such models, the internal spaces are direct products of the aforementioned minimal conformal theories, which constitute the basic ``building blocks'', yielding a total central charge for the corresponding Liouville (non-critical-string) model of the form $C=22 + \delta c$. The quantity $\delta c$ is the so-called central charge deficit in the language of non-critical (Liouville) strings. The authors of \cite{Antoniadis:1988aa} argued that the fact that the central charges of all  the building-block conformal field theories are $\ge \frac{1}{2}$, implies that each member of the infinite family of ``internal'' spaces contains only a finite number of building-block conformal theories, and as such, the central charge deficit $\delta c$  can only take on discrete values. 
Given that each of these discrete ``internal'' spaces corresponds, according to the approach of \cite{Antoniadis:1988aa}, to a given cosmology, one arrives at the conclusion that, in such a construction, there is only a discrete set of values that the Hubble expansion rate of the Universe can take, leading to a discrete series of inflationary vacua, as in our case above. 
At present, a formal connection of our approach with such non-critical discrete Liouville cosmologies is not derived, but we thought it would be interesting to mention it as another case of realization of a discrete series of de-Sitter spacetimes.

Since, as mentioned above, the case $\widetilde{\lambda} \geq 0$, $\gamma < 0$, with $\abs{\gamma/\delta} \le 1 $, 
which includes this discrete series of vacua (see fig.~\ref{figure0}), does not include the inflationary scenario of \cite{Dorlis:2024yqw}, this series of de-Sitter vacua cannot be smoothly embedded in that framework. Nonetheless, there is the possibility that such a series arises at a post-inflationary era of the StRVM model \cite{Dorlis:2024yqw,bms,bms2,ms1,ms2}. This can happen, for instance, due to an, as yet, unknown microscopic mechanism of generation of such a series of de-Sitter vacua, through some kind of phase transition, by means of which a post-inflationary StRVM universe finds itself in one of these vacua. From then on, it evolves, through a series of tunneling effects, so that  a Minkowski spacetime is reached asymptotically ({\it cf. }figure \ref{figure0}). It is currently unknown whether such a mechanism can explain the observed accelerated expansion of the Universe,  
corresponding to the present-era value of the cosmological constant~\cite{Planck},
 but we think that such speculative scenarios are worthy of further exploration. 

\section{Dynamical System for the case \texorpdfstring{$\widetilde{\lambda} < 0$ and $\gamma <0$}{lambda0 < 0 and gamma < 0} and its stability analysis}\label{sec:five}
\begin{figure}[ht!]
    \centering
\includegraphics[width=0.495\textwidth]{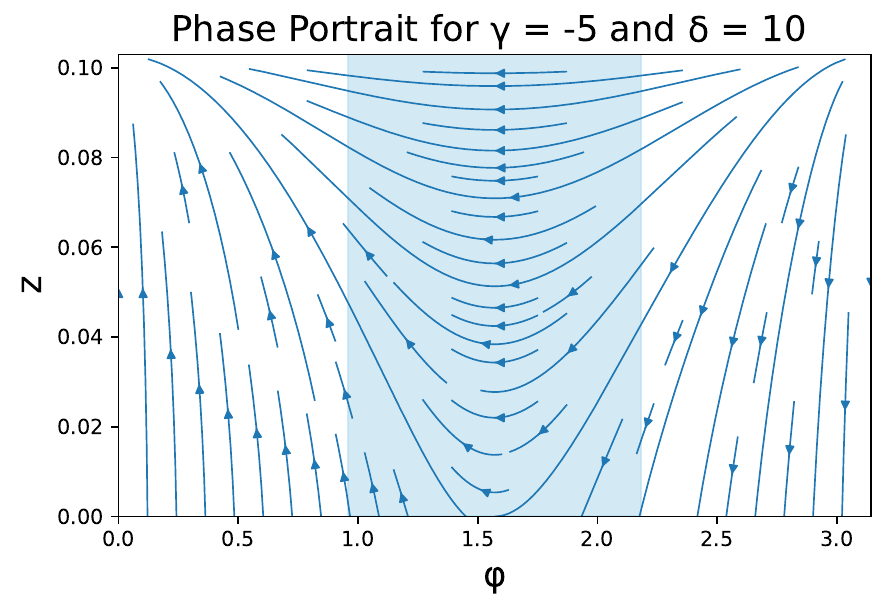}\hfil\includegraphics[width=0.505\textwidth]{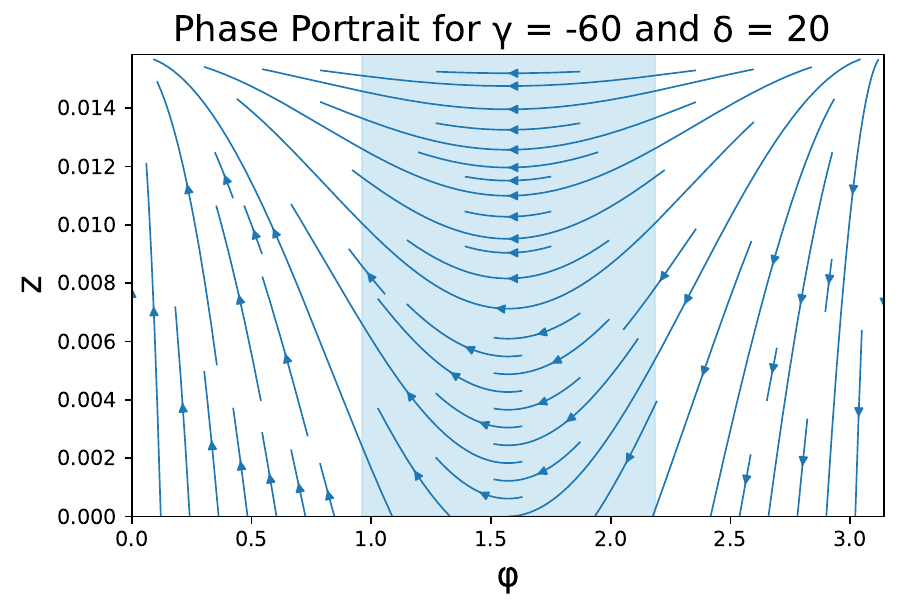}
    \caption{Phase space of \eqref{ode55} for $\abs{\gamma/\delta} \leq 1$ (left panel) and for $\abs{\gamma/\delta} \geq 1$ (right panel). }
    \label{figure11}
\end{figure}

In the previous section we established a lower bound \eqref{lowerbound}, for the ratio $\abs{\gamma/\delta}$, in order to have a de-Sitter spacetime. Now, in this case we must have\,:
\begin{align}\label{53}
 V>0 \Rightarrow   \gamma\, z + \cos \left( \delta\,z\right) >0\,,\,\,\text{with}\,\,\,\, z\equiv \kappa\,b\,.
\end{align}
Since $\gamma<0$ and $z>0$, it follows that \eqref{53} is true when and only when $\cos \left( \delta\,z\right) >0$, implying $\delta\,z < \pi/2$, and $z$ variable is bounded and belongs to $z \in [0,r)$ where $r$ is the root of the equation $V=0$ or equivalently $\gamma\, z + \cos \left( \delta\,z\right) =0\,.$
This equation does not have analytical solution\footnote{Essentially it involves the inverse function of spherical Bessel function $y_0 (x)$~\cite{abramowitz1965handbook}, which in general has not yet been studied in the literature, at least to our knowledge.}, but for every choice of $\gamma\,,\delta$ has a solution. The first solution that corresponds to $V=0$ is $r$, as our analysis no longer holds after this point. Thus, we have the dynamical system\,:
\begin{align}
\begin{aligned}\label{ode55}
& \varphi^{\prime}=\left[3 \cos \varphi-\sqrt{\frac{3}{2}}\,\delta \,\left(\frac{\frac{\delta}{\gamma}\,{\rm sin}(\delta\,z)-1}{\frac{\delta}{\gamma}\,{\rm cos}(\delta\,z)+ \delta\,z}\right)\right] \sin \varphi \\
& z^{\prime}=\sqrt{6} \cos \varphi\,,
\end{aligned}
\end{align}
where $\varphi \in[0, \pi]$, $z \in[0,r)$. To find the critical points \eqref{fp} of \eqref{ode55}, we notice that from $z^{\prime}$ equation implies $\varphi=\pi/2$ and hence from $\varphi^{\prime}$ equation we obtain\,:
\begin{align}
    \frac{\delta}{\gamma}\,{\rm sin}(\delta\,z)-1 =0 \Rightarrow z_1=\frac{\pi- \operatorname{arcsin}\left(\frac{\gamma }{\delta }\right) +2 \pi  c_1}{\delta } \ \ \text{and} \ \ z_2=\frac{ \operatorname{arcsin}\left(\frac{\gamma }{\delta }\right)+2 \pi  c_2}{\delta }\,,
\end{align}
with $c_{1,2} \in \mathbb{Z}$ and $\abs{\gamma/\delta} \leq 1$. The, $c_{1,2}$ will be restricted in view of $z_{1,2} \in [0,r)$\,:
\begin{align}
    0 \leq z_1 < r < \frac{\pi}{2 \delta} \,\stackrel{(\gamma=-\delta)}{\Rightarrow} \,0 \leq 3+ 4 c_1 < \frac{2 \,\delta\,r}{\pi} < 1 \Rightarrow -0.75 \leq c_1 < -0.5 \,.
\end{align}
Since, as  $c_{1} \in \mathbb{Z}$ there is no critical points for this solution for the system \eqref{ode55}. Similarly we have that\,:
\begin{align}
    0 \leq z_2 < r < \frac{\pi}{2 \delta} \,\stackrel{(\gamma=-\delta)}{\Rightarrow} \,0 \leq  4 c_2-1 < \frac{2 \,\delta\,r}{\pi} < 1 \Rightarrow 0.25 \leq c_2 < 0.5 \,,
\end{align}
so again we have no solution and consequently the dynamical system \eqref{ode55} does not have fixed points. We depict the relevant phase space in figure \ref{figure11}.

\section{String/dbrane inspired potential for compactification axions}\label{sec:6}
\begin{figure}[htbp!]
    \centering
\includegraphics[width=0.515\textwidth]{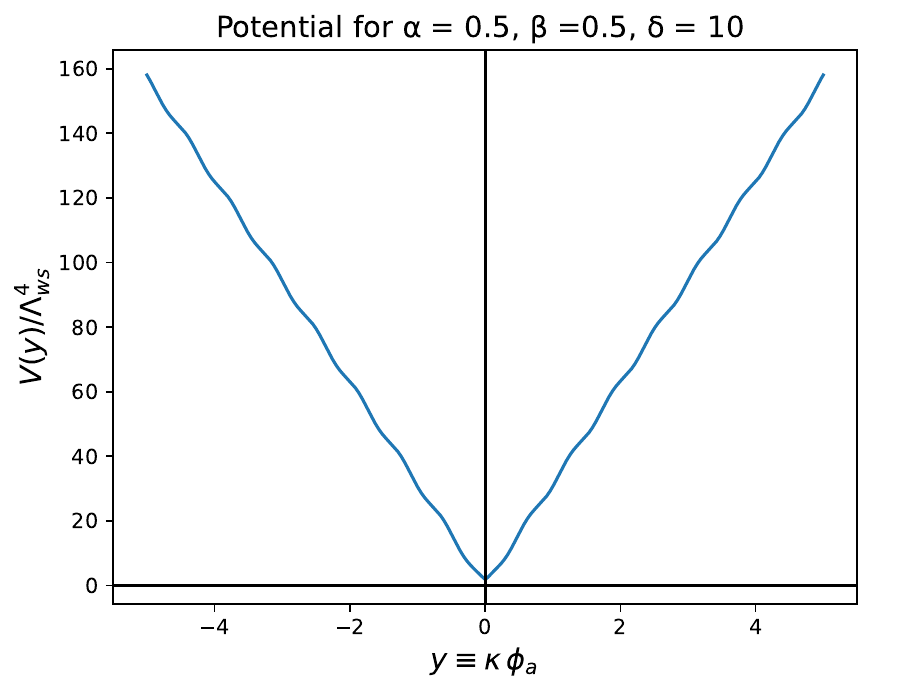}\hfil\includegraphics[width=0.485\textwidth]{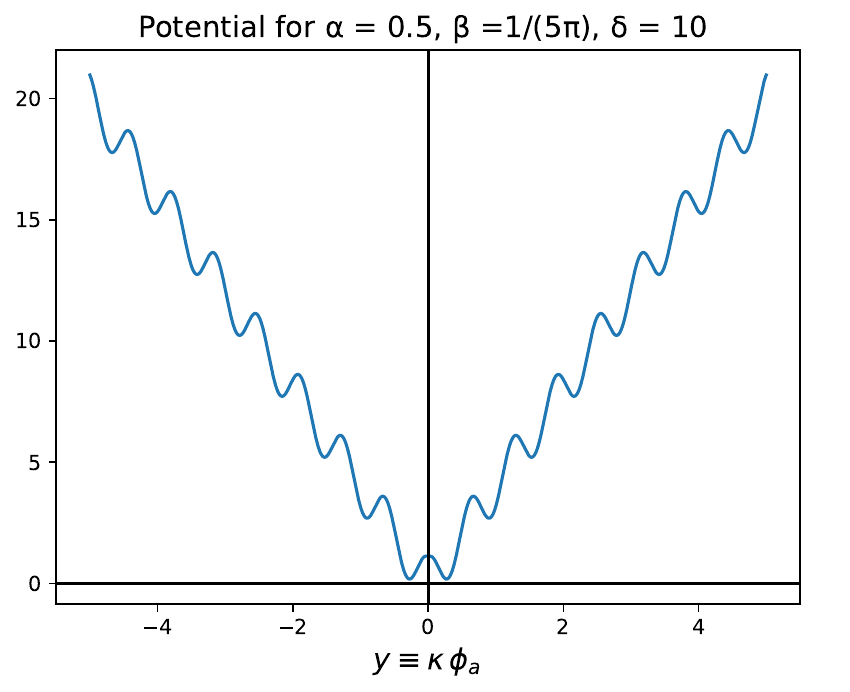}
    \caption{Potential \eqref{Vphistring} for different choice of parameters which has immediate effect on phase space and the relevant cosmologies. Note that the parameter $\beta$ is responsible for the appearance or not of the infinite series of critical points in the potential, where $\beta=1/2\pi$ corresponds the bifurcation point.  \underline{Left panel}: $\beta>1/ 2\pi$. \underline{Right panel}: $\beta<1/ 2\pi$. }
    \label{figure22}
\end{figure}

In this final section we use the dynamical-system formalism to study the following potential, arising in string theory scenarios (like IIB string model,
involving D5-branes), as advocated in \cite{silver}, and discussed also in the introduction of the current work\footnote{In this study, we omit terms proportional to $\phi_a \cos \left(\phi_a / 2 \pi f_a\right)$, as they are subdominant in the slow roll parameters \cite{silver}.} ({\it cf.} \eqref{Vmonodr2}):
\begin{equation}\label{Vphistring}
V(\phi_a)=\Lambda_2^4\, \ell^2\sqrt{1 + \Big(\frac{\phi_a(x)}{\ell^2\, f_a}\Big)^2}+\Lambda_{\rm ws}^4 \cos \left(\frac{\phi_a}{2 \pi f_{a}} \right)\,,
\end{equation}
where $\phi_a= a \,f_a $ has mass dimension $+1$\, and denotes a dominant compactification-axion species, and $\Lambda_2^4$ given in \eqref{Vlincomp}. The potential \eqref{Vphistring} is plotted in figure \ref{figure22} for two indicative choices of parameters, and we have made the definitions\,:
\begin{align}\label{defs}
\alpha = 1/ \ell^2\,, \quad \beta =\left( \Lambda_2/{\Lambda_{\rm ws}}\right)^4 ,\, \quad \delta=1/(2 \pi\,\kappa\,f_a)\,, \quad y=\kappa\,\phi_a\,,
\end{align}
which will be used in what follows.
Following the analysis presented in section \ref{sec:dynsys}, we have for the potential \eqref{Vphistring}\,:
\begin{align}\label{lamdadef}
    \lambda=-\frac{V_{, \phi_a}}{\kappa V}  = \frac{\alpha  \delta  \left(\sin (\delta  y)-\frac{4 \pi ^2 \alpha  \beta  \delta  y}{\sqrt{4 \pi ^2 \alpha ^2 \delta ^2 y^2+1}}\right)}{\beta  \sqrt{4 \pi ^2 \alpha ^2 \delta ^2 y^2+1}+\alpha  \cos (\delta  y)}\,,
\end{align}
where we used \eqref{defs}. 
We also have that\,:
\begin{align}
    1-\Gamma &=1+\frac{\left(\beta  \sqrt{4 \pi ^2 \alpha ^2 \delta ^2 y^2+1}+\alpha  \cos (\delta  y)\right) \left(\left(4 \pi ^2 \alpha ^2 \delta ^2 y^2+1\right)^{3/2} \cos (\delta  y)-4 \pi ^2 \alpha  \beta \right)}{\alpha  \sqrt{4 \pi ^2 \alpha ^2 \delta ^2 y^2+1} \left(\sqrt{4 \pi ^2 \alpha ^2 \delta ^2 y^2+1} \sin (\delta  y)-4 \pi ^2 \alpha  \beta  \delta  y\right)^2} \nonumber\\
     &= 1 + \frac{\alpha  \delta ^2}{\lambda^2} \,\frac{ \left(\left(4 \pi ^2 \alpha ^2 \delta ^2 y^2+1\right)^{3/2} \cos (\delta  y)-4 \pi ^2 \alpha  \beta \right)}{\left(4 \pi ^2 \alpha ^2 \delta ^2 y^2+1\right)^{3/2} \left(\beta  \sqrt{4 \pi ^2 \alpha ^2 \delta ^2 y^2+1}+\alpha  \cos (\delta  y)\right)}\,.
\end{align}
Thereby, the dynamical system in this scenario is \,:
\begin{align}
\begin{aligned}\label{61}
& x^{\prime}=(x^2-1)\left[3x-\sqrt{\frac{3}{2}} \,\lambda \right] \\
& \lambda^{\prime}=\sqrt{6}\,x \left(\lambda^2 +  \frac{\alpha  \delta ^2 \left(\left(4 \pi ^2 \alpha ^2 \delta ^2 y^2+1\right)^{3/2} \cos (\delta  y)-4 \pi ^2 \alpha  \beta \right)}{\left(4 \pi ^2 \alpha ^2 \delta ^2 y^2+1\right)^{3/2} \left(\beta  \sqrt{4 \pi ^2 \alpha ^2 \delta ^2 y^2+1}+\alpha  \cos (\delta  y)\right)}\right)\,.
\end{aligned}
\end{align}
Some remarks are in order at this point, regarding the system \eqref{61}. Firstly, under the simultaneous transformations $x \rightarrow -x,\, y \rightarrow-y$ (with the latter corresponding to $\lambda \rightarrow -\lambda$, on account of \eqref{lamdadef}), the system is invariant. Therefore, without loss of generality we can assume that $y\geq0$ or $y\leq0$. Moreover, as the transformation $\delta \rightarrow - \delta$ does not alter our system, we may assume, that $\delta \geq 0$. In such a case, in order to maintain the transplanckian-censorship conjecture~\cite{trans}, we must have $\delta \gtrsim 1/(2 \pi) \simeq 0.159$. In addition, in the limit $y \rightarrow \pm \infty$, the system reduces to that of\,\cite{Dorlis:2024yqw}. We define the $\widetilde{\lambda}$ variable and then we compactify it, as follows\,:
\begin{align}
\widetilde{\lambda} =-\frac{1}{y} = - \frac{1}{\kappa\,\phi_a}\,,\qquad   \zeta=\frac{\widetilde{\lambda}}{\widetilde{\lambda}+1}\quad \Rightarrow \quad \widetilde{\lambda}=\frac{\zeta}{1-\zeta}\,,
\end{align}
which takes values\,in\,the range $\zeta \in[0,1)$, for $y \in(-\infty,0]$. On setting  $x=\cos \varphi$, where $\varphi \in[0, \pi]$, we can write the system \eqref{61} in the form\,:

\begin{align}
\begin{aligned}\label{dynsysstring}
\varphi^{\prime} & = \left[3 \cos \varphi+\sqrt{\frac{3}{2}}\,\,\frac{\alpha  \delta  \left(\frac{4 \pi ^2 \alpha  \beta  \delta  (\zeta -1)}{\zeta  \sqrt{\frac{4 \pi ^2 \alpha ^2 \delta ^2 (\zeta -1)^2}{\zeta ^2}+1}}+\sin \left(\frac{\left(1-\zeta\right)}{\zeta} \,\delta \right)\right)}{\beta  \sqrt{\frac{4 \pi ^2 \alpha ^2 \delta ^2 (\zeta -1)^2}{\zeta ^2}+1}+\alpha  \cos \left(\frac{\left(1-\zeta\right)}{\zeta} \,\delta \right)} \,\right] \,\sin \varphi\, \\
      \zeta^{\prime}  &= \sqrt{6}\, \zeta^2 \,\cos \varphi\,.
\end{aligned}
\end{align}
Below we restrict ourselves to indicative cases of the parameters with physical significance. We recall that $l \geq 1$, hence $0<\alpha \leq 1$, and we pick, as representative values, $\alpha \ll 1$ and $\alpha =0.5$, corresponding to very large and medium-range (string scale) compactification radii in our string system, respectively.

\begin{figure}[ht!]
    \centering
\includegraphics[width=0.8\textwidth]{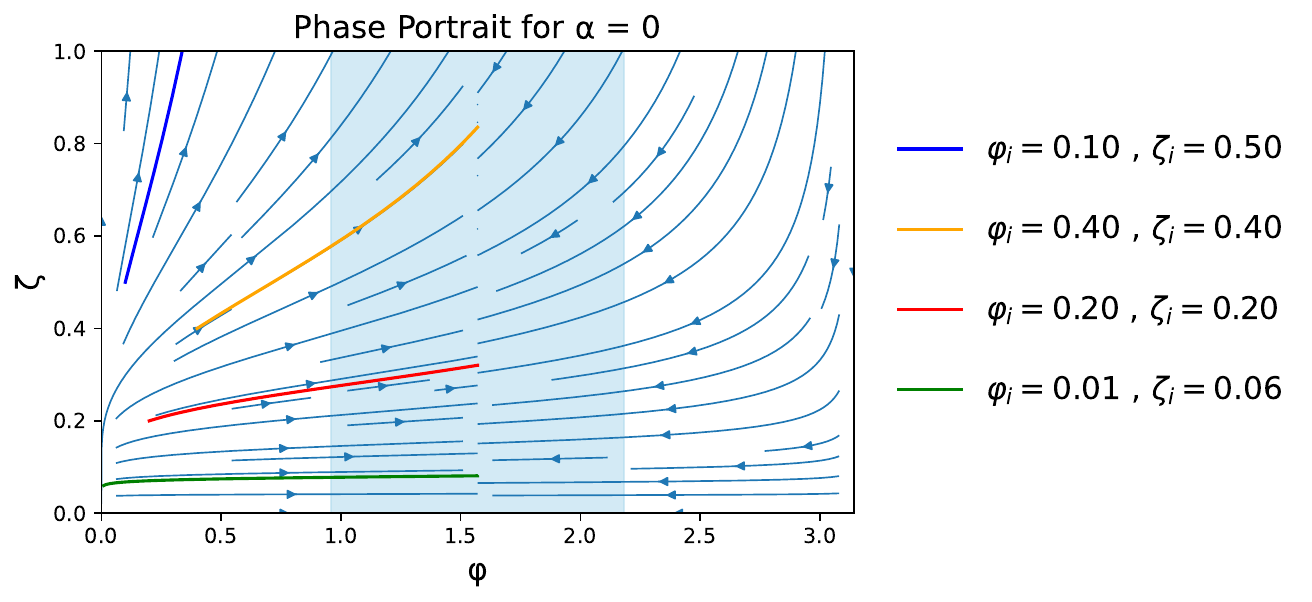}
    \caption{Phase space of dynamical system \eqref{dynsysstringlim} for different initial conditions. Not all initial conditions result in eternal de-Sitter-like spacetimes, like the blue curve corresponding to the initial conditions $\varphi_i =0.10$, $\zeta_i =0.50$.}
    \label{figure14}
\end{figure}

We first note that, in the formal case where $\alpha =0$ (infinite compactification radius),
the dynamical system \eqref{dynsysstring} assumes a particularly simple form, independent of the parameters $\beta$ and $\delta$\,:
\begin{align}
\begin{aligned}\label{dynsysstringlim}
\varphi^{\prime} & = 3 \cos \varphi\, \sin \varphi\ \\
      \zeta^{\prime}  &= \sqrt{6}\,\zeta^2\,\cos \varphi\,.
\end{aligned}
\end{align}
This form is to be understood from the approximately de Sitter form of the potential \eqref{Vphistring}\,:
\begin{align}
 \lim_{\ell \gg 1}V(\phi_a) \simeq \Lambda_2^4\, \ell^2 \,,   
\end{align}
which is independent of the scale $\Lambda_{\rm ws}$, given that the latter accompanies the subleading cosine term in \eqref{Vphistring}. Thus, this case leads only to eternal de-Sitter in view of the above. We have the critical points $(\varphi,\zeta)$ to be \,${\rm O_2}=(0,0)\, ,\ {\rm A_2}=(\pi/2 ,0)\, ,\ {\rm B_2}=(\pi ,0)\,,\ {\rm E}_{\zeta}=(\pi/2 ,\zeta)$ with $\zeta \in (0,1)$, thus we need to find the stability of these points, although the aforesaid argument is valid. The Jacobian matrix of this system is given by\,:
\begin{equation}
   J( \varphi, \zeta) =
\begin{bmatrix}
   3 \cos (2 \varphi ) & 0 \\[1.5ex]
     -\sqrt{6} \,\zeta ^2 \sin (\varphi ) & 2 \sqrt{6} \,\zeta  \cos (\varphi ) 
\end{bmatrix}
\end{equation}
and very easily we find that for both fixed points ${\rm O_2}=(0,0)\, ,{\rm B_2}=(\pi ,0)$ we have\,:
\begin{equation}
   J( 0, 0) =
\begin{bmatrix}
   3 & 0 \\[1.5ex]
    0 & 0
\end{bmatrix}
= J( \pi, 0) \,\,\,\text{with eigenvalues} \,\,\left\{3,0\right\}\,.
\end{equation}
Thence they are unstable points, because the positive eigenvalues guarantee that there is at least one unstable direction\,\cite{B_hmer_2016}. Now for the other two points (more accurately the $E_{\zeta}$ corresponds to a family of critical points), both have eigenvalues $\left\{-3,0\right\}$ which means that  we need to employ other methods, like center manifold theory in order to determine their stability properties. We present the analysis in subsection \ref{subsecC2} of Appendix \ref{app3} and find, as expected, that the fixed points ${\rm A_2}(\varphi=\pi/2,\zeta=0)$  and ${{\rm E}_{\zeta}}(\varphi=\pi/2,\zeta)$ with $\zeta \in (0,1)$ correspond to stable-node critical points. The phase space of this limiting case is portrayed in figure \ref{figure14}.

\begin{figure}[htbp!]
    \centering
\includegraphics[width=0.563\textwidth]{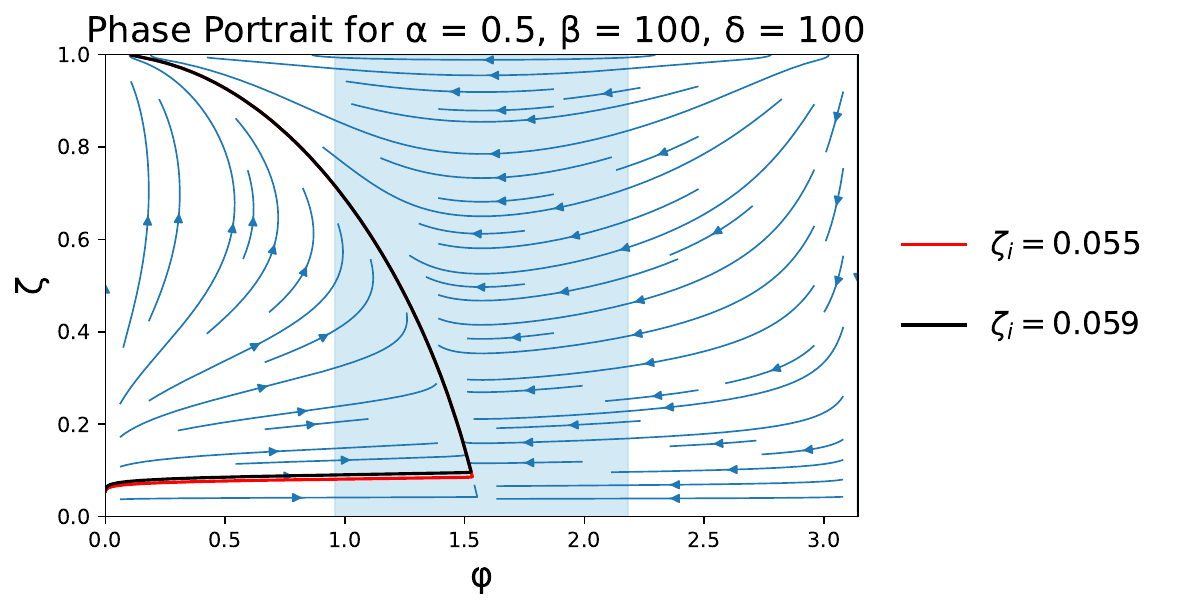}\hfil\includegraphics[width=0.437\textwidth]{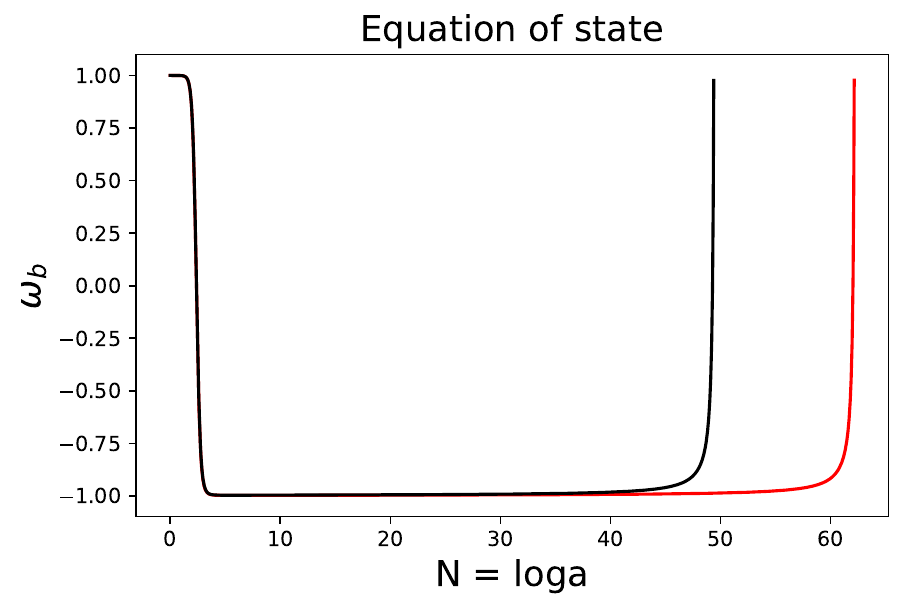}
    \caption{Phase space (left panel) and EoS (right panel) for $\varphi_i =0.001$ and values of $\zeta_i$ for which we obtain ${\mathcal N}_{\rm e} = \mathcal O(50-60)$.}
    \label{figure15}
\end{figure}

Now, we make some general comments regarding the dynamical system \eqref{dynsysstring}. First of all, we observe that the critical points $ {\rm O_3}=(0,0)\, ,\ {\rm A_3}=(\pi/2 ,0)\,,\text{and}\,\ {\rm B_3}=(\pi ,0)$  essentially capture the behavior at infinity, {\it i.e.} when $y \rightarrow -\infty \Rightarrow \zeta \rightarrow 0$. This limit, as previously mentioned, corresponds to a system whose potential is basically a linear one and its well studied, so for the different values of $\alpha\,,\beta\,,\delta$ (excluding\, $\alpha=0$) the stability of those points are known and they are unstable, saddle and unstable respectively\,\cite{Dorlis:2024yqw}. The question then arises as to whether there exist additional critical points and what are the physical implications of those points in string cosmology. We examine the physically interesting case~\cite{silver} in which the worldsheet instanton effects are suppressed compared to the rest of the term in the potential \eqref{Vphistring}: $\alpha = 0.5$, $\beta =100$, $\delta =100$. This system exhibits only the aforementioned fixed points and no more. One is able to find initial conditions that result in metastable inflation with the phenomenologically desired e-foldings~\cite{Planck}, as shown in figure \ref{figure15}. There are suppressed modulations, as in the case of the periodically modulated StRVM~\cite{Dorlis:2024uei}. The fact that we have a finite lifetime in the inflationary era, is consistent with swampland criteria~\cite{swamp1,swamp2,swamp3,swamp4,swamp5}, as expected from a microscopic quantum string cosmology model.

We can show that the dynamical system \eqref{dynsysstring} will have extra critical points when $\varphi = \pi/2$ and when the following equation is satisfied, i.e.\,:
\begin{align}\label{74}
  \frac{4 \pi ^2 \alpha  \beta  \delta  (\zeta -1)}{\zeta  \sqrt{\frac{4 \pi ^2 \alpha ^2 \delta ^2 (\zeta -1)^2}{\zeta ^2}+1}}+\sin \left(\frac{\left(1-\zeta\right)}{\zeta} \,\delta \right) = 0\,,
\end{align}
which gives a bound $\beta \leq 1/ (2 \pi ) \approx 0.159$, by exploiting the relation $\abs{\sin x} \leq 1$.
However, since the $\Lambda_{\rm ws}$ scale corresponds to worldsheet instantons, which are exponentially suppressed compared to the $\Lambda_2$ scale ({\it i.e.} $\beta \gg 1$), these cases are not physical. If we pick the parameters to fall in this range (for instance, $\alpha = 0.5$, $\beta =1/(5 \pi)$, $\delta =10$) stable spiral\,/\,node, saddle points arise in an analogous situation to that in subsection \ref{sec4.2} for the potential \eqref{pot}. 

\begin{figure}[ht!]
    \centering
    \includegraphics[width=\textwidth]{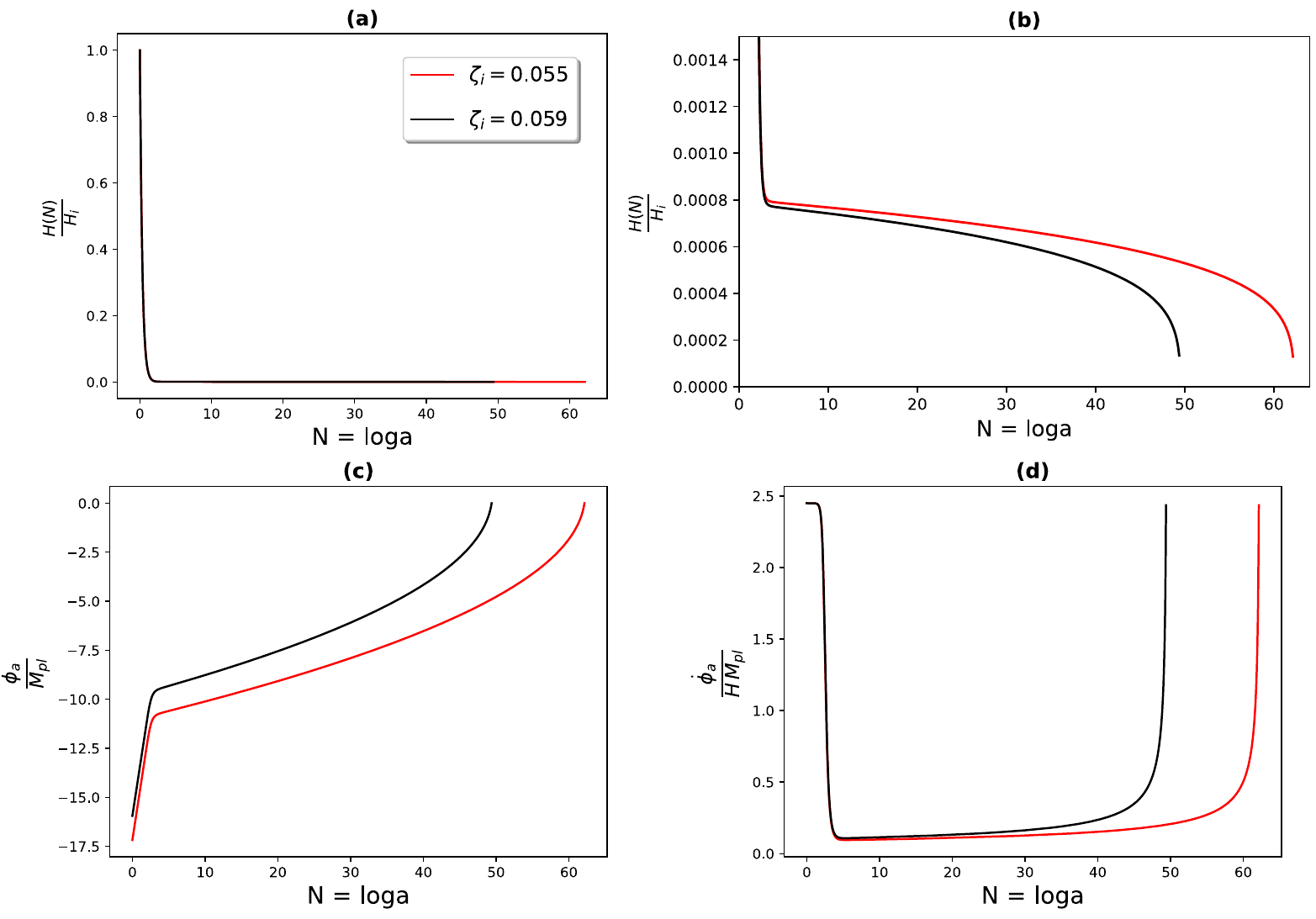}
    \caption{The plots correspond to the string/brane-inspired case of figure \ref{figure15}, for the initial conditions $\varphi_i =0.001$, and two characteristic values of the $\zeta_i$ parameter, as shown in the upper left panel (a). The upper panels (a), (b) show the change of the Hubble rate relative to an initial value $H_i$, whilst the lower panels (c), (d) depict the compactification axion $\phi_a$ and its cosmic rate ${\dot \phi}_a$, demonstrating again the satisfaction of the slow-roll conditions during inflation, which, depending on the value of $\zeta_i$, can last as long as 60 e-foldings, consistent with the data~\cite{Planck}.}
    \label{fig:new14}
\end{figure}

Now, regarding the slow-roll conditions in this string/dbrane cosmological scenario, we have that (similarly to eqs.\eqref{Hratio} in subsection \ref{sec4.2})\,:
\begin{align}
    \frac{H(N)}{H_i}= \sqrt{\frac{\frac{\beta}{\alpha} \,\sqrt{1+4\,\pi^2 \,\alpha^2\,\delta^2\,\left(1-1/\zeta     \right)^2} + \cos \left(\delta\,\left(1-1/\zeta     \right) \right)}{\frac{\beta}{\alpha} \,\sqrt{1+4\,\pi^2 \,\alpha^2\,\delta^2\,\left(1-1/\zeta_i     \right)^2} + \cos \left(\delta\,\left(1-1/\zeta_i    \right) \right)}}\,\frac{\sin\varphi_i}{\sin\varphi}\,,
\end{align}
and \eqref{dbb} holds as it is. We depict the relevant results in figure \ref{fig:new14}, following the discussion in subsection \ref{sec4.2}. The validity of 
the slow-roll conditions for inflation in this string/brane-inspired model, from a dynamical-system viewpoint, becomes apparent from the various panels in the figure. This, together with the pertinent results for the cases of subsection \ref{sec4.2}, indicates that one can in fact obtain an inflationary state, via a slow-roll approximation, in diverse settings that agree with the plethora of the available phenomenological/observational data~\cite{Planck}.
The (world-sheet-instanton-induced) periodic modulations of the potential \eqref{Vphistring} ensure a closer agreement of the values of the slow-roll inflationary parameters $\epsilon_i, i=1,2,3$, and the spectral index $n_s$~\cite{inflationaris,Planck} with the corresponding values inferred from the cosmological measurements, as in the case of \cite{Dorlis:2024uei}.

\section{Conclusions and Outlook}\label{sec:7}

In this work we have performed a dynamical-system study of three-parameter inflationary potentials \eqref{pot} of axion-like fields, with periodic modulations over linear terms, both breaking explicitly the axion-shift symmetry. Such kind of potentials also characterize brane-compactification induced ones, which are included in the study, corresponding to some regions of the parameter space. Using a suitable set of parameters
$\gamma$ and $\delta$, \eqref{gd}, and $\widetilde{\lambda} $, \eqref{lambda0}, we have performed a classification of the initial conditions and the parameter-space regimes leading to stable or metastable de-Sitter (inflationary) vacua, appearing as fixed points of the corresponding flows in the appropriate phase diagrams. We have found several cases for which it is possible to select suitable values for the potential parameters and initial conditions so that the phenomenologically relevant  duration of inflation of order ${\mathcal O}(50-60)$ e-foldings can be achieved, during which the slow-roll conditions \eqref{Hdot}, ensuring an approximately constant $H_I$, can be satisfied. The periodic modulations of the inflationary potentials discussed in this work, then, will ensure a closer agreement of the theoretically-predicted values of the slow-roll inflationary parameters $\epsilon_i, \,i=1,2,3$ and the running spectral index $n_s$ with the corresponding values inferred from cosmological measurements~\cite{Planck}, in similar spirit to the case studied in \cite{Dorlis:2024uei}.

The case of dominant linear-terms has already been studied in a previous work by (some of) the authors~\cite{Dorlis:2024yqw}, where the (approximately) linear terms are induced by condensates of anomalous (3+1)-dimensional gravitational Chern-Simons terms coupled to the axion. In this work we have extended the analysis to regions of the parameter space for which the periodic modulations of the linear terms are of comparable size or stronger than the linear terms in the axion potentials and classified the relevant initial conditions in the dynamical-system parameters which lead to inflationary eras with phenomenologically relevant duration. 
An interesting result in this region of the parameter space concerns the case of a discrete (infinite) sequence of fixed points corresponding to classically stable de-Sitter vacua, with 
decreasing values of the local minima (as the axion-field values increase). 
This sequence resembles, superficially at least, the situation of discrete inflation encountered in minimal non-critical string-theory (Liouville) models. 
The dynamical system analysis does not explain the exit mechanism from such vacua, due to their stability. Nonetheless we conjecture that once the physical system enters one of these vacua, due, {\it e.g.}, to some phase transitions of the underlying microscopic model, then non-perturbatively-induced tunneling effects may take over, thus allowing a sequence of transitions of the system to lower-energy vacua, until the system reaches asymptotically a zero-potential-minimum, Minkowski spacetime or a small effective cosmological constant. The lifetimes of the various de-Sitter vacua depend on the regime of the pertinent parameters. The physical significance of such a phase is still not clear. The only thing that is clear is that this is a distinct phase from the running-vacuum linear-axion condensate inflation. 
At this stage, one cannot exclude the possibility that such a phase may even be related to the current-era of acceleration of the universe. This is definitely an issue we plan to put research emphasis on in the immediate future.

Last but not least, we have applied the dynamical system analysis to study the brane-compactification-induced single-axion-monodromy inflationary scenario. We have demonstrated that, there too, there is compatibility with phenomenologically-relevant inflationary scenarios, by a suitable choice of the parameter space region.

We hope that our work will constitute a useful addition to the interesting research map of the axion-monodromy inflationary models. The latter, being primarily associated at a microscopic level with string models, involve more than a single axion-like field, the string-model-independent one and the axions coming from compactification~\cite{svrcek}. These kinds of axions come with different coupling constants. 
The easiest (although complicated enough, due to the increased number of parameters involved) scenario to apply our dynamical-system analysis to is models with two axions, both characterized by linear potentials  
with periodic modulations, like the ones appearing in \cite{Zhou:2020kkf,Mavromatos:2022yql}.
Such models differ by the hierarchies of the energy scales characterizing  the linear and periodic modulation axion-potential terms. Both categories exhibit a rich phenomenology, as far as the spectrum of primordial black holes, and consequently the profiles of GWs during the early stages of radiation, are concerned. In fact, as shown in \cite{Mavromatos:2022yql}, there are distinguishing features in the GW profiles between these two categories of models, 
which are in principle detectable in future interferometric devices, such as LISA or others~\cite{Berti:2005ys,Crowder:2005nr}.

\appendix

\acknowledgments
The work of P.D. and M.V. is supported by a graduate scholarship from the
National Technical University of Athens (Greece). The work of N.E.M. is supported in part by UK Engineering and Physical Sciences Research Council (EPSRC) under the research grant  ST/X000753/1. The work of S.-N.V. is supported
by the Hellenic Foundation for Research and Innovation (H.F.R.I. (EL.ID.EK.)) under the ``5th Call for H.F.R.I. Scholarships to PhD Candidates" (Scholarship Number: 20572).N.E.M. acknowledges participation in the COST Association Actions CA21136 “Addressing observational tensions in cosmology with systematics and fundamental physics (CosmoVerse)” and CA23130 "Bridging high and low energies in search of quantum gravity (BridgeQG)".

\section*{APPENDICES} 
\section[\appendixname~\thesection]{Purely Cosine axion potential}\label{app2}
\begin{figure}[htbp!]
    \centering
\includegraphics[width=0.51 \textwidth]{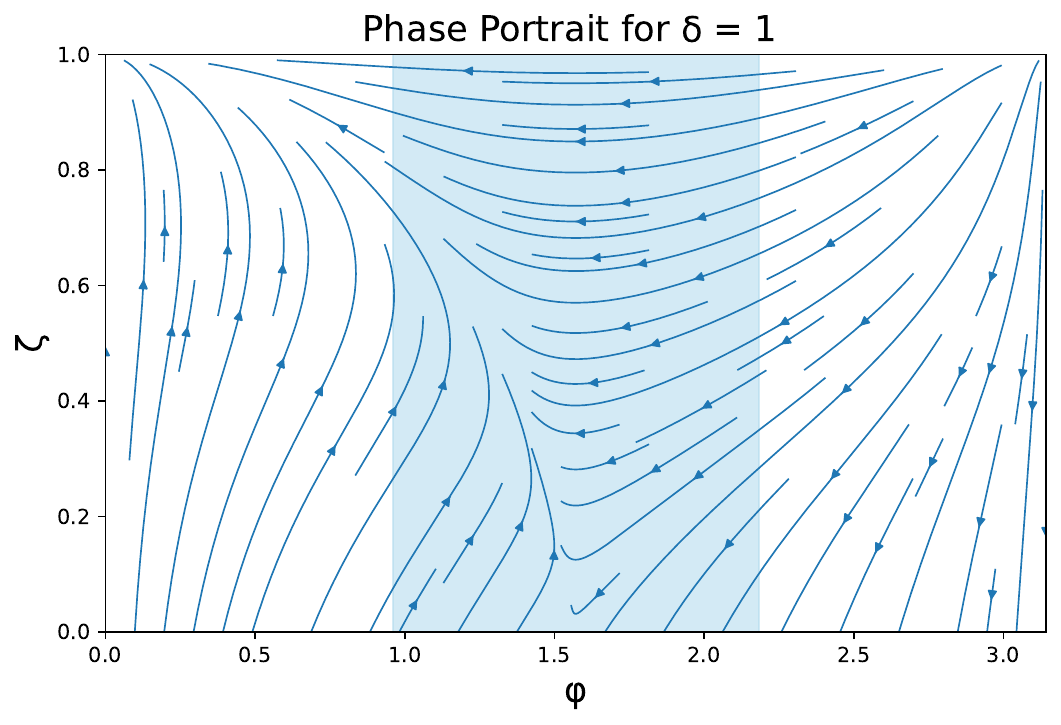}\hfil\includegraphics[width=0.49\textwidth]{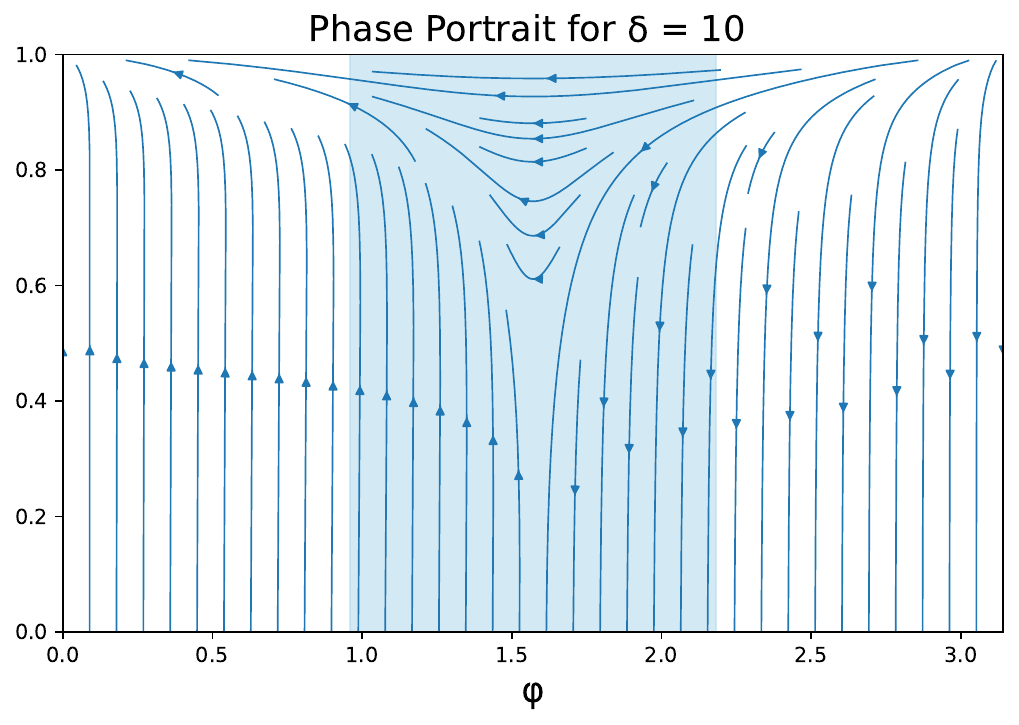}
    \caption{Phase space for purely cosine case for two values of the $\delta$ parameter.}
    \label{figurepc}
\end{figure}

In this appendix, we briefly study the case when we have only the cosine potential:
\begin{align}
V(b) = \Lambda_1^4 \, {\rm cos}\left(\frac{b}{f_b}\right)\,.
\end{align}
Since the above potential is periodic, we are able to study the dynamics of the axion field in just one period\,\cite{Hossain:2023lxs}. We have the relations\,:
\begin{align}
V_{, b} (b) =  - \frac{ \Lambda_1^4}{f_b} \, {\rm sin}\left(\frac{b}{f_b}\right), \ \ \ \ \ V_{, b b} (b) =  - \frac{ \Lambda_1^4}{f_b ^2} \, {\rm cos}\left(\frac{b}{f_b}\right) \,.
\end{align}
The procedure is exactly the same as in section \ref{sec:dynsys}, where after introducing the EN variables, the $\lambda$ variable takes the form\,:
\begin{align}
    \lambda=-\frac{V_{, b}}{\kappa V}  = \frac{1}{\kappa\,f_b} \, {\rm tan} \left(\frac{b}{f_b}\right) \ \ \text{and} \ \ \Gamma = \frac{V V_{, b b}}{V_{, b}^2}=- \frac{1}{\kappa^2 \,f^2_b\,\lambda^2}\,.
\end{align}
Hence, we obtain the dynamical system\,:
\begin{align}
\begin{aligned}\label{A4}
& x^{\prime}=(x^2-1)\left[3x-\sqrt{\frac{3}{2}} \,\lambda \right] \\
& \lambda^{\prime}=\sqrt{6}\,x\,(\lambda^2 +\delta^2 ) \,,
\end{aligned}
\end{align}
with $\delta \equiv 1/(\kappa\,f_b) $\,. Of course, since the transformation $\delta \rightarrow - \delta$, does not change the dynamical system \eqref{A4}, without loss of generality we assume that $\delta > 0$. Now, we set $x=\cos \varphi$, where $\varphi \in[0, \pi]$ and then the system \eqref{A4} takes the form\,:
\begin{align}
\begin{aligned}\label{A5}
    \varphi^{\prime} & = \left[3 \cos \varphi-\sqrt{\frac{3}{2}}\, \lambda \,\right] \,\sin \varphi, \\
     \lambda^{\prime}  &= \sqrt{6}\, \cos \varphi\,(\lambda^2 +\delta^2 ) \,.
     \end{aligned}
\end{align}
In principle, $\lambda$ can take on any value on the real axis, and the dynamical system \eqref{A5} is invariant under the simultaneous transformations ( $\varphi \rightarrow \pi-\varphi, \,\lambda\rightarrow-\lambda$ ), implying that we can assume $\lambda \geq 0$, without loss of generality. The phase space is bounded through the following change of variable\,:
\begin{align}
\zeta=\frac{\lambda}{\lambda+1} \Rightarrow \lambda=\frac{\zeta}{1-\zeta}\,,
\end{align}
which takes on values\,in\,the region $\zeta \in[0,1)$, for $\lambda \in[0,+\infty)$. Also, note that\,:
\begin{align}
     \zeta^{\prime}= \lambda^{\prime}(\zeta)\,\frac{d\zeta}{d\lambda}=\lambda^{\prime}(\zeta)\,\frac{1}{(1+\lambda(\zeta))^2} =  \sqrt{6}\, \cos \varphi\,\frac{(\lambda^2(\zeta) +\delta^2 )}{(1+\lambda(\zeta))^2} =  \sqrt{6}\, \cos \varphi\,(\zeta^2 +\delta^2 \,(1-\zeta)^2)\,.\nonumber
\end{align}
Thence, the system of ODE \eqref{A5} reads\,:
\begin{align}
\begin{aligned}\label{54}
    \varphi^{\prime} & = \left[3 \cos \varphi-\sqrt{\frac{3}{2}}\, \frac{\zeta}{1-\zeta} \,\right] \,\sin \varphi, \\
     \zeta^{\prime}  &= \sqrt{6}\, \cos \varphi\,(\zeta^2 +\delta^2 \,(1-\zeta)^2)\,.
     \end{aligned}
\end{align}
Now, we have the following critical point of \eqref{54}: from the $\zeta^{\prime}=0$ equation, we have $\varphi=\pi/2 $, and for this choice we substitute in the equation for $ \varphi^{\prime} =0$, and then we obtain the critical point $ {\rm A_4}=(\pi/2,0)$. The Jacobian stability matrix for the system \eqref{54} is\,:
\begin{align}\label{A8}
   J( \varphi, \zeta) =
\begin{bmatrix}
 \frac{\sqrt{\frac{3}{2}} \zeta  \cos \varphi }{\zeta -1}+3 \cos (2 \varphi ) & -\frac{\sqrt{\frac{3}{2}} \sin \varphi}{(\zeta -1)^2} \\
 -\sqrt{6} \left(\delta ^2 (\zeta -1)^2+\zeta ^2\right) \sin \varphi  & 2 \sqrt{6} \left(\delta ^2 (\zeta -1)+\zeta \right) \cos \varphi  \\
\end{bmatrix}
\end{align}
We now proceed in finding the eigenvalues of the Jacobian matrix at the critical point. For the point ${\rm A_4}$, we substitute into the Jacobian \eqref{A8} the values $\varphi =\pi/2, \,\zeta=0$, thus obtaining:
\begin{align}
J(\pi/2,0)=\begin{bmatrix}
-3 & -\sqrt{\frac{3}{2}} \\
-\sqrt{6} \,\delta^2 & 0
\end{bmatrix}\,,
\end{align}
which has eigenvalues $\left\{\frac{1}{2}\left(-3-\sqrt{3} \sqrt{3+4 \delta^2}\right), \frac{1}{2}\left(-3+\sqrt{3} \sqrt{3+4 \delta^2}\right)\right\}$. As the eigenvalues come with opposite sign  and equal magnitude, this is a saddle point and a hyperbolic one (as neither of the eigenvalues have zero real part). This implies that some trajectories will be repelled, while others will be attracted\,\cite{B_hmer_2016}. The pertinent cosmologies are characterized by the respective initial conditions. Some indicative phase portraits,  for different values of $\delta$, are shown in fig.~\ref{figurepc}.

\section[\appendixname~\thesection]{Eigenvalues corresponding to ${\rm C_{\,c_1}}(\gamma, \delta)$ and ${\rm D_{c_2}}(\gamma, \delta)$}
\label{appB}
In this appendix, we provide the complete analysis of the eigenvalues that correspond to the Jacobian (stability) matrix of the two families ${\rm C_{\,c_1}}(\gamma, \delta)$ and ${\rm D_{c_2}}(\gamma, \delta)$ appearing in the case when $\abs{\gamma/\delta }\leq 1 $ in subsection \ref{sec:3.1}. The eigenvalues in each critical point can easily be computed from the corresponding Jacobian, and in general have the form\,\cite{B_hmer_2016}:
\begin{equation}
\lambda_{1,2}=\frac{1}{2}\left( \dfrac{\partial \varphi^{\prime}}{\partial \varphi}+ \dfrac{\partial \zeta^{\prime}}{\partial \zeta}\right) \pm \frac{1}{2} \sqrt{\left( \dfrac{\partial \varphi^{\prime}}{\partial \varphi}- \dfrac{\partial \zeta^{\prime}}{\partial \zeta}\right)^2+4 \, \dfrac{\partial \varphi^{\prime}}{\partial \zeta}  \dfrac{\partial \zeta^{\prime}}{\partial \varphi}}\,.
\end{equation}
For the first family ${\rm C_{\,c_1}}(\gamma, \delta)$ the corresponding eigenvalues are given by\,:
\begin{align}
\lambda_{1,2}= \frac{1}{2} \Bigg[ -3 \pm \sqrt{\frac{12 \delta ^2 \left(\gamma  \delta  \sqrt{1-\frac{\gamma ^2}{\delta ^2}} \operatorname{arcsin}\left(\frac{\gamma }{\delta }\right)+\delta ^2-\gamma ^2+\pi  (2 c_1+1) \gamma  \delta  \sqrt{1-\frac{\gamma ^2}{\delta ^2}}\right)}{\left(\delta  \sqrt{1-\frac{\gamma ^2}{\delta ^2}}+\pi  (2 c_1 \gamma +\gamma )+\gamma  \operatorname{arcsin}\left(\frac{\gamma }{\delta }\right)\right)^2}+9}\Bigg]\nonumber\,.
\end{align}
We observe that the quantity in the square root, call it $f_1(\gamma,\delta,{c_1})$, is not always positive due to the last term in the numerator of the fraction. Thus, one can distinguish two cases: one in which $f_1(\gamma,\delta,{c_1}) <0$, and another with $f_1(\gamma,\delta,{c_1}) \geq 0$. In the former case, the fixed points are stable
spiral, which means that the solutions approach the critical points as $N \rightarrow \infty$, but not from a definite direction\,\cite{boehmer2010} and the eigenvalues have the form $\lambda_1=\alpha+i \beta$, $\lambda_2=\alpha-i \beta$ with $\alpha<0$ and $\beta \neq 0$. In fact, we can determine the $c_1$ corresponding to these fixed points, because the $f_1(\gamma,\delta,{c_1})$ is quadratic to this parameter, and by standard analysis we obtain\,:
\begin{align}
    f_1(\gamma,\delta,{c_1})<0 \Rightarrow 0 \leq c_1 \leq \left \lfloor \frac{1}{2 \pi} \left(\frac{\left(4 \delta ^2+3\right)}{3}  \sqrt{\frac{\delta ^2}{\gamma ^2} -1}-\operatorname{arcsin}\left(\frac{\gamma }{\delta }\right)-\pi \right) \right \rfloor\,,
\end{align}
where $\lfloor \ldots \rfloor$ denotes the floor function and we took into account equation \eqref{49}. An equivalent form reads\,:
\begin{align}
    \sqrt{\frac{\delta ^2}{\gamma ^2}-1}-\operatorname{arcsin}\left(\frac{\gamma }{\delta }\right)-\pi <0\,.
\end{align}
The case in which $f_1(\gamma,\delta,{c_1}) \geq 0$ is attained when \,:
\begin{align}
    c_1 \geq \left \lceil \frac{1}{2 \pi} \left(\frac{\left(4 \delta ^2+3\right)}{3}  \sqrt{\frac{\delta ^2}{\gamma ^2} -1}-\operatorname{arcsin}\left(\frac{\gamma }{\delta }\right)-\pi \right) \right \rceil\,,
\end{align}
where $\lceil \ldots \rceil$ is the ceiling function, and we note that $c_1 \in  \mathbb{N}\cup\{0\}$. Moreover, as the following inequality holds\,:
\begin{align}
  \abs{- 3 } > \sqrt{f_1(\gamma,\delta,{c_1})}\,,
\end{align}
the fixed points correspond to stable-node points,  that is, every orbit tends to the critical point in a
definite direction as $N \rightarrow \infty$ \cite{boehmer2010} and $\lambda_1<0, \lambda_2<0$. 

Similarly, for the second family ${\rm D_{c_2}}(\gamma, \delta)$ we have the eigenvalues:
\begin{align}
\lambda_{1,2}= \frac{1}{2} \Bigg[ -3 \pm \sqrt{\frac{12 \delta ^2 \left(\gamma  \delta  \sqrt{1-\frac{\gamma ^2}{\delta ^2}} \operatorname{arcsin}\left(\frac{\gamma }{\delta }\right)+\delta ^2-\gamma ^2-2 \pi  c_2 \gamma  \delta  \sqrt{1-\frac{\gamma ^2}{\delta ^2}}\right)}{\left(\delta  \sqrt{1-\frac{\gamma ^2}{\delta ^2}}-2 \pi  c_2 \gamma +\gamma  \operatorname{arcsin}\left(\frac{\gamma }{\delta }\right)\right)^2}+9}\Bigg]\nonumber\,.
\end{align}
In this case,  the quantity inside the square root, call it $f_2(\gamma,\delta,{c_2})$, is manifestly positive~\footnote{This follows from the fact 
that $\gamma<0$, $\abs{\gamma/\delta} \leq 1$, as well as the use of the identity $\operatorname{arcsin}\left(-x\right)=-\operatorname{arcsin}\left(x\right)$, with  $\operatorname{arcsin}\left(x\right)>0$ for $x \in(0, 1]$, with $\delta^2 -\gamma^2 >0$.}. In addition, since the following inequality holds\,:
\begin{align}
  \abs{- 3 } < \sqrt{f_2(\gamma,\delta,{c_2})}\,,
\end{align}
we have that the fixed points of this family are saddle, since $\lambda_1>0$ and $\lambda_2<0$.

\section[\appendixname~\thesection]{Center manifold theory}
\label{app3}
In this appendix, we use center manifold theory~\cite{Bahamonde:2017ize,boehmer2010,Boehmer:2011tp} in order to study the stability properties of two special cases of dynamical systems, where linear stability analysis fails.
\subsection[Appendix C.1]{Stability analysis for $\abs{\gamma / \delta} = 1$ case}\label{subsec1AppC}
First, we examine the case when $\abs{\gamma/\delta }= 1 $, whereupon the dynamical system~\eqref{ode2} becomes\,:
\begin{align}\label{odeC}
\begin{aligned}
\varphi^{\prime} & = \left[3 \cos \varphi-\sqrt{\frac{3}{2}}\,\left( \frac{\zeta - \zeta\, {\rm sin}\left(\frac{\left(1-\zeta\right)}{\zeta} \,\delta \right)}{1-\zeta +\zeta\,\frac{1}{\delta}\, {\rm cos}\left(\frac{\left(1-\zeta\right)}{\zeta} \,\delta \right)} \right) \,\right] \,\sin \varphi\, \\
      \zeta^{\prime}  &= \sqrt{6}\, \cos \varphi\,\zeta^2 \,.
\end{aligned}
\end{align}
Except for the fixed points at infinity ({\it i.e.} $\zeta=0$), the system \eqref{odeC} possesses also one family of critical points. Basically, what happens in this case is that the two families of subsection~\ref{sec:3.1} degenerate into one. Eq. \eqref{twofam}, then, takes the form\,:
\begin{align}\label{newfam}
\zeta_3 =\frac{2\delta}{\pi+2\delta+4 \pi c_1} \ \ \  \text { with }\,\, c_{1} \in \mathbb{N}\cup\{0\}\,.
\end{align}
The eigenvalues $\lambda_i, \,i=1,2$ of the Jacobian for this family, ${\rm E_{\,c_1}}( \delta)=\left(\pi/2,\,\zeta_3 \right)$\,, can be computed as 
$\lambda_{1,2}= -3,0$. Hence, the linear stability theory fails to provide the stability properties for this family. To study the stability of this case, therefore, we employ center manifold theory~\cite{Bahamonde:2017ize,boehmer2010,Boehmer:2011tp}. In this approach, we first select randomly~\footnote{The reader should notice that any point of this family can be treated equivalently, given that $c_1$,\,$\delta$ are free parameters.} a point from the ${\rm E_{\,c_1}}( \delta)$ family, by choosing specific values for $c_1$,\,$\delta$ and 
"move" it to the origin via a coordinate transformation:  $\varphi \to \theta + \frac{\pi}{2}$ and $\zeta \to \psi + \zeta_3$. In the new coordinates, $\theta,\psi$, the system \eqref{odeC} becomes: 
\begin{align}\label{odeC2}
\begin{aligned}
\theta^{\prime} & = \left[-3 \sin \theta-\sqrt{\frac{3}{2}}\,\left( \frac{1 - {\rm sin}\left(\frac{\left(1-\psi - \zeta_3\right)}{\psi + \zeta_3} \,\delta \right)}{\frac{1}{\psi + \zeta_3}-1 +\frac{1}{\delta}\, {\rm cos}\left(\frac{\left(1-\psi - \zeta_3\right)}{\psi + \zeta_3} \,\delta  \right)} \right) \,\right] \,\cos \theta\, \\
      \psi^{\prime}  &= -\sqrt{6}\, \sin \theta\,(\psi + \zeta_3)^2 \,.
\end{aligned}
\end{align}
The Jacobian matrix of the system \eqref{odeC2} at the origin reads: 
\begin{equation}\label{A19}
    \left. J \right|^{\theta=0}_{\psi=0} =\begin{bmatrix}
        -3 & 0\\
        -\sqrt{6} \,\zeta_3^2 & 0
        \end{bmatrix}\,,
\end{equation}
with the eigenvalues and the corresponding eigenvectors given by:
\begin{align}
&\lambda_1=0 \ \to \ \mathbf{u}=\left(\begin{array}{c}
 0 \\
 1 \\
\end{array}\right)\,,
\\
 &\lambda_2=-3 \ \to \ \mathbf{v}=\left(\begin{array}{c}
 \frac{\sqrt{6}}{2\,\zeta_3^2} \\
 1 \\
\end{array}\right)\,.
\end{align}
The eigenvalue $\lambda_1 = 0$ and the eigenvector $\mathbf{u}$ span the center subspace for the fixed point, while the eigenvalue $\lambda_2 = -3$ and the eigenvector $\mathbf{v}$ span the stable subspace. Now, we find new variables which diagonalize the Jacobian matrix \eqref{A19} and as such we have the linearized system\,:
\begin{equation}
    \left(\begin{array}{c}
 \theta^{\prime} \\
 \psi^\prime \\
\end{array}\right)=  \left. J \right|^{\theta=0}_{\psi=0} \left(\begin{array}{c}
 \theta \\
 \psi \\
\end{array}\right)\,,
\label{linear_theta_z}
\end{equation}
and the Jacobian can be written\,: 
\begin{align*}
     \left. J \right|^{\theta=0}_{\psi=0}=P D P^{-1} \,,
\end{align*}
where\,:
\begin{equation}  
    P =\begin{bmatrix}
       0  & \frac{\sqrt{6}}{2\,\zeta_3^2}\\
        1 & 1
        \end{bmatrix} , \ P^{-1} =\begin{bmatrix}
       - \frac{2\,\zeta_3^2}{\sqrt{6}} &  1\\
        \frac{2\,\zeta_3^2}{\sqrt{6}} & 0
        \end{bmatrix}  , \  D =\begin{bmatrix}
        0 &  0\\
        0 & -3
        \end{bmatrix} 
\end{equation}
which diagonalize the Jacobian matrix \eqref{A19} at the origin. Following the analysis presented in appendix B of~\cite{Dorlis:2024yqw}, we multiply \eqref{linear_theta_z} with $P^{-1}$ from the left on both sides and obtain:
\begin{equation}
    P^{-1} \left(\begin{array}{c}
 \theta^{\prime} \\
 \psi^\prime \\
\end{array}\right)= D P^{-1} \left(\begin{array}{c}
 \theta \\
 \psi \\
\end{array}\right)\,.
\label{diagonilise_coords}
\end{equation}
Therefore, a suitable coordinate transformation is\,:
\begin{equation}
     \left(\begin{array}{c}
 U \\
 V \\
\end{array}\right)=P^{-1} \left(\begin{array}{c}
 \theta \\
 \psi \\
\end{array}\right)=\begin{bmatrix}
       - \frac{2\,\zeta_3^2}{\sqrt{6}} &  1\\
        \frac{2\,\zeta_3^2}{\sqrt{6}} & 0
        \end{bmatrix} \left(\begin{array}{c}
 \theta \\
 \psi \\
\end{array}\right)\,,
\label{coord_U_V}
\end{equation}
which produces\,:
\begin{equation}
    U=- \frac{2\,\zeta_3^2}{\sqrt{6}}\,\theta +\psi \ \  \text{and} \ \  V=\frac{2\,\zeta_3^2}{\sqrt{6}}\,\theta\,.
\end{equation}
In addition, eq. \eqref{diagonilise_coords} assumes the form\,: 
\begin{equation}
   \left(\begin{array}{c}
 U^\prime \\
 V^\prime \\
\end{array}\right) = D \left(\begin{array}{c}
 U \\
 V \\
\end{array}\right)=\begin{bmatrix}
        0 &  0\\
        0 & -3
        \end{bmatrix} \left(\begin{array}{c}
 U \\
 V \\
\end{array}\right)\,\Rightarrow  U^\prime = 0\,,
    \,V^\prime = -3 V\,.
\end{equation}
Hence, we can express the dynamical system \eqref{odeC2} in terms of the the new variables $U$ and $V$ as\,:
\begin{align}\label{A28}
\begin{aligned}
   U^\prime &= A U + f(U,V) \,,\\  
   V^\prime &= B V + g(U,V)\,,
   \end{aligned}
\end{align}
where $A=0,B=-3$ and the functions $f,g$ are given by:
\begin{align}
    f(U,V) &= \left[\sqrt{6} \sin \left(\frac{\sqrt{6}}{2\,\zeta_3^2}\,V\right)+ \frac{1 - {\rm sin}\left(\frac{\left(1-U-V - \zeta_3\right)}{U+V + \zeta_3} \,\delta \right)}{\frac{1}{U+V + \zeta_3}-1 +\frac{1}{\delta}\, {\rm cos}\left(\frac{\left(1-U-V - \zeta_3\right)}{U+V + \zeta_3} \,\delta  \right)} \,\right] \zeta_3^2\,\cos \left(\frac{\sqrt{6}}{2\,\zeta_3^2}\,V\right) \nonumber\\ 
  & -\sqrt{6}\, \sin \left(\frac{\sqrt{6}}{2\,\zeta_3^2}\,V\right)\,(U+V + \zeta_3)^2  \label{f(u_v)}\\
   g(U,V) &= -\left[\sqrt{6} \sin \left(\frac{\sqrt{6}}{2\,\zeta_3^2}\,V\right)+ \frac{1 - {\rm sin}\left(\frac{\left(1-U-V - \zeta_3\right)}{U+V + \zeta_3} \,\delta \right)}{\frac{1}{U+V + \zeta_3}-1 +\frac{1}{\delta}\, {\rm cos}\left(\frac{\left(1-U-V - \zeta_3\right)}{U+V + \zeta_3} \,\delta  \right)} \,\right] \zeta_3^2\,\cos \left(\frac{\sqrt{6}}{2\,\zeta_3^2}\,V\right) \nonumber \\
    &+3 V \label{g(u_v)}\,.
\end{align}
We next apply the Local Center Manifold Theorem~\cite{perko2001}. According to this theorem, we assume the following expansion for the function $h(U)$ of our center manifold\,: 
\begin{align}\label{uexp}
h(U)=a_1 U^2+a_2 U^3+\ldots\,, 
\end{align} and, thus, we need to solve the differential equation, with $V=h(U)$ valid locally \ :
\begin{align}\label{A32}
    &\frac{dh}{dU}\, \left[A U + f(U,h(U)) \right] - B h(U) - g(U,h(U))=0 \,.
\end{align}
We can solve eq.~\eqref{A32} for the first terms in the polynomial expansion \eqref{uexp} of  $h(U)$, keeping only the leading orders and using \eqref{newfam}. We obtain\,:
\begin{equation}
    h(U)=-\frac{\delta \,(\pi+2\delta+4 \pi c_1)^2}{12 (\pi+4 \pi c_1 )} U^2+\frac{(\pi+2\delta+4 \pi c_1 )^3 \left(\pi  (4 c_1+1) \left(\delta ^2+3\right)+2 \delta ^3\right)}{36 \,(4 \pi c_1+\pi )^2} U^3 +\mathcal{O}(U^4)\,.
\end{equation}
Having determined $h(U)$, we are now in position to define the stability properties of our dynamical system \eqref{odeC2} by simply studying the stability of the equation\,:
\begin{align}\label{A34}
   u^{\prime}(N) = A \,u(N) + f(u(N),h(u(N)))\,.
\end{align}
To this end, we make a series expansion of the right-hand-side of \eqref{A34}, to obtain\,:
\begin{align}\label{A35}
    u^{\prime}(N) = \frac{\delta \, (\pi+2\delta+4 \pi c_1)^2}{4 (4 \pi  c_1+\pi )}\, u(N)^2 -\frac{\delta ^2 \,(\pi+2\delta+4 \pi c_1)^4}{24 (4 \pi  c_1+\pi )^2}\,u(N)^3 +\mathcal{O}\left(u(N)^4\right)\,.
\end{align}
All the information about the stability of the direction related to the zero eigenvalue of the fixed point $(\varphi=\frac{\pi}{2},\,\zeta_3)$ is now encoded in the sign of the constant in front of the term $u(N)^2$. Since it is manifestly positive, we have an instability along the eigenvector of the zero eigenvalue. Consequently, the ${\rm E_{\,c_1}}( \delta)=\left(\pi/2,\,\zeta_3 \right)$ family corresponds to saddle critical points.
\subsection[Appendix C.2]{Stability analysis for the $\alpha=0$ (infinite-compactification-radius) case of the string/brane-inspired model of \cite{silver}}\label{subsecC2}
To study this case, we use center manifold theory~\cite{Bahamonde:2017ize,boehmer2010,Boehmer:2011tp} upon focusing on the ${\rm A_2}=(\pi/2 ,0)$ fixed point of the dynamical system \eqref{dynsysstringlim}. The procedure is the same as before. We "move" the point, via a coordinate transformation, to the origin, $\varphi \to \theta + \frac{\pi}{2}$, and in the new coordinates $\theta,\zeta$, the system \eqref{dynsysstringlim} becomes: 
\begin{align} 
\begin{aligned} \label{thetaode}
    \theta^{\prime}& =-3\cos \theta \,\sin \theta\,,
    \\
    \zeta^{\prime}& = -\sqrt{6}\,\zeta^2 \sin \theta \,,
\end{aligned}
\end{align}
where now the fixed point ${\rm A_2}(\varphi=\pi/2 ,\zeta=0)$ amounts to ${\rm A_2}(\theta=0,\zeta=0)$. The Jacobian matrix of the system \eqref{thetaode} reads: 
\begin{equation}
    J =\begin{bmatrix}
        -3 \cos (2 \theta ) & 0 \\
        -\sqrt{6} \zeta^2 \cos (\theta ) &  -2 \sqrt{6} \zeta \sin (\theta )
        \end{bmatrix} \quad \Rightarrow \quad  \left. J \right|^{\theta=0}_{\zeta=0} =\begin{bmatrix}
        -3 & 0\\
        0 & 0
        \end{bmatrix}\,,
    \label{Jacobian2}
\end{equation}
with the eigenvalues and the corresponding eigenvectors given by:
\begin{align}
 &\lambda_1=-3 \ \to \ \mathbf{u}=\left(\begin{array}{c}
 1 \\
 0 \\
\end{array}\right)\,,
\\
 &\lambda_2=0 \ \to \ \mathbf{v}=\left(\begin{array}{c}
 0 \\
 1 \\
\end{array}\right)\,.
\end{align}
The eigenvalue $\lambda_1 = -3$ and the eigenvector $\mathbf{u}$ span the stable subspace for the fixed point, while the eigenvalue $\lambda_2 = 0$ and the eigenvector $\mathbf{v}$ span the center subspace. The Jacobian can be written in the form\,: 
\begin{align*}
     \left. J \right|^{\theta=0}_{\zeta=0}=P D P^{-1} \,,
\end{align*}
where\,:
\begin{equation}  
    P =\begin{bmatrix}
       0  & 1\\
        1 & 0
        \end{bmatrix} , \ P^{-1} =\begin{bmatrix}
       0 &  1\\
        1 & 0
        \end{bmatrix}  , \  D =\begin{bmatrix}
        0 &  0\\
        0 & -3
        \end{bmatrix} \,.
\end{equation}
Moreover we have\,:
\begin{equation}
    P^{-1} \left(\begin{array}{c}
 \theta^{\prime} \\
 \zeta^\prime \\
\end{array}\right)= D P^{-1} \left(\begin{array}{c}
 \theta \\
 \zeta \\
\end{array}\right) \quad \Rightarrow  \quad \left(\begin{array}{c}
 U \\
 V \\
\end{array}\right)=P^{-1} \left(\begin{array}{c}
 \theta \\
 \zeta \\
\end{array}\right)=\begin{bmatrix}
       0 &  1\\
        1 & 0
        \end{bmatrix} \left(\begin{array}{c}
 \theta \\
 \zeta \\
\end{array}\right)\,,
\label{diagonilise_coords1}
\end{equation}
and as such\,:
\begin{equation}
    U= \zeta \ \  \text{and} \ \  V=\theta\,.
\end{equation}
Now, eq. \eqref{diagonilise_coords1} reads\,: 
\begin{equation}
   \left(\begin{array}{c}
 U^\prime \\
 V^\prime \\
\end{array}\right) = D \left(\begin{array}{c}
 U \\
 V \\
\end{array}\right)=\begin{bmatrix}
        0 &  0\\
        0 & -3
        \end{bmatrix} \left(\begin{array}{c}
 U \\
 V \\
\end{array}\right)\,\Rightarrow  U^\prime = 0\,,
    \,V^\prime = -3 V\,.
\end{equation}
In this scenario, \eqref{A28} has $A=0,B=-3$ and the functions $f,g$ are\,:
\begin{align}
    f(U,V) &= - \sqrt{6}\, U^2 \cos V \nonumber \\
    g(U,V) &= 3V - 3 \cos V \sin V\,.
\end{align}
By simply applying the Local Center Manifold Theorem \cite{perko2001}, we assume the $h(U)$ expansion for the function  of our center manifold, as in \eqref{uexp} and, on setting $V=h(U)$, we solve \eqref{A32}, to obtain\,:
\begin{equation}
    h(U)=0+\mathcal{O}(U^5)\,,
\end{equation}
which means that the center manifold coincides with the $U$ axis. With this result, the stability properties of our dynamical system \eqref{thetaode} are determined by \eqref{A34}, thus\,:
\begin{align}\label{A341}
   u^{\prime}(N) = A \,u(N) + f(u(N),h(u(N))) \Rightarrow u^{\prime}(N) = - \sqrt{6}\, u(N)^2\,,
\end{align}
implying that the critical point ${\rm A_2}(\theta=0,\zeta=0)$ is a stable node. By the same token, we find that the family ${{\rm E}_{\zeta}}(\theta=0,\zeta)$, with $\zeta \in (0,1)$, also corresponds to a family of stable-node critical points.

\bibliography{DMVV_DYN}

\end{document}